\documentclass[reprint, rmp, aps, nofootinbib, twocolumn]{revtex4-1}
\usepackage{calc}
\usepackage{amsmath, amssymb}
\usepackage{amsbsy}
\usepackage{graphicx}

\def\Xint#1{\mathchoice
{\XXint\displaystyle\textstyle{#1}}%
{\XXint\textstyle\scriptstyle{#1}}%
{\XXint\scriptstyle\scriptscriptstyle{#1}}%
{\XXint\scriptscriptstyle\scriptscriptstyle{#1}}%
\!\int}
\def\XXint#1#2#3{{\setbox0=\hbox{$#1{#2#3}{\int}$}
\vcenter{\hbox{$#2#3$}}\kern-.5\wd0}}

\def\dashint{\Xint-}
\DeclareMathOperator{\tr}{Tr}                   
\def\abs#1{\left \vert #1 \right \vert}
\def\mean#1{\left \langle #1 \right \rangle}

\begin{document}
\title{Physics of the Riemann Hypothesis}

\author{D{\'{a}}niel Schumayer}
\email{dschumayer@physics.otago.ac.nz}
\author{David A. W. Hutchinson}
\affiliation{Jack Dodd Centre for Quantum Technology, Department of Physics,
University of Otago, Dunedin, New Zealand}

\begin{abstract}
Physicists become acquainted with special functions early in their studies.
Consider our perennial model, the harmonic oscillator, for which we need
Hermite functions, or the Laguerre functions in quantum mechanics. Here we
choose a particular number theoretical function, the Riemann zeta function
and examine its influence in the realm of physics and also how physics may
be suggestive for the resolution of one of mathematics' most famous unconfirmed
conjectures, the Riemann Hypothesis. Does physics hold an essential key to
the solution for this more than hundred-year-old problem? In this work we
examine numerous models from different branches of physics, from classical
mechanics to statistical physics, where this function plays an integral
role. We also see how this function is related to quantum chaos and how its
pole-structure encodes when particles can undergo Bose-Einstein condensation
at low temperature. Throughout these examinations we highlight how physics
can perhaps shed light on the Riemann Hypothesis. Naturally, our aim could
not be to be comprehensive, rather we focus on the major models and aim to
give an informed starting point for the interested Reader.
\end{abstract}

\date{}
\maketitle
\tableofcontents

\section{\label{sec:intro}Introduction}

\hfill \parbox{80mm}{\raggedleft
                     {\emph{`Can you do Addition?' the White Queen asked.
                            `What's one and one and one and one and one
                             and one and one and one and one and one?' \\
                             `I don't know,' said Alice. `I lost count.'}}\\
                     (Lewis Carroll - Through the Looking Glass)
                     } \\
\vspace*{3mm}

Counting, in the broadest sense, is probably the oldest mathematical activity
and not even uniquely ours. Even animals can distinguish one, two and three,
maybe just by recognising a pattern, but only humans have developed an abstract
language, mathematics or more specifically number theory, which accurately
describes the properties of numbers.

In the following we will focus on the border between physics and number theory,
and more precisely, how the Riemann-zeta function, $\zeta(s)$, appears in quite
different areas of physics. This review does not intend to be comprehensive,
rather would like to offer a panoramic view and give a feeling as to why many
physicist find beauty in the structure of this seemingly random function and
what one might learn from it. We collect examples from diverse realms of physics,
from classical mechanics to condensed matter physics, where the Riemann-zeta
function or its `descendants' play a significant role. Due to space limitations
we do not aspire to be mathematically precise in our derivations, but we give
physical arguments to support results, and also direct the Reader to relevant
sources.

\section{\label{sec:HistoryAndNecessities}
         Historical background and `Mathematical necessities'}

\hfill \parbox{42mm}{\raggedleft
                     {\emph{God invented the integers; \\ all else is the work of man.}}
                     (Leopold  Kronecker)
                     } \\
\vspace*{3mm}

Natural numbers form the basis of our arithmetic, with various operations
defined among these numbers. All of us learn to use four basic operations:
addition, subtraction, multiplication and division. The latter, division,
hides one of the most enigmatic internal structures of the set of the natural
numbers, namely that there are special numbers, the {\emph{primes}}, among the
natural numbers which cannot be divided by any other natural number, other
than unity and themselves, without a remainder. Euclid of Alexandria proved
that there are infinitely many such numbers. Later, Eratosthenes of Cyrene
gave a theoretical algorithm, a sieve, for finding these primes amongst the
natural numbers. Despite all efforts in the last two thousand
years, the efficient determination as to whether a given number is prime
or not still proves a remarkable challenge.

It is not hard to understand why the distribution of primes could captivate the
imagination of many mathematicians and physicists. These numbers seem to obey
two contradictory principles. Firstly, they seem to appear randomly among
composite numbers, but secondly they also appear to obey strict rules governing
their distribution.


Apart from Euclid's, numerous proofs exist for the infinitude of
the prime numbers \cite{Ribenboim1991}. Euler, at the early age of
30, proved a stronger statement \cite{Euler1737},
\begin{equation}
   \sum_{p \ {\mathrm{prime}} }{\frac{1}{p}} = \infty.
\end{equation}
This formula clearly proves Euclid's statement but it also demonstrates
the frequent occurrence of prime numbers amongst composite numbers. A natural
continuation of his work was the analysis of the arithmetic properties of the
series, $\sum{n^{-k}}$. Substituting $k=1$ into this expression we recover the
well-known, divergent harmonic series. Conversely, if $k > 1$ the summation
converges. Euler also showed~\cite{Euler1737} -- using the fundamental theorem
of arithmetic -- that this series can be written as an infinite product over
the prime numbers, $p$, such that
\begin{equation}
   \label{eq:EulerProduct}
   \zeta(k) =
   \sum_{n=1}^{\infty}{\frac{1}{n^{k}} } 
   =
   \prod_{p}{\left ( 1- \frac{1}{p^{k}}\right )^{-1}}.
\end{equation}
One may interpret through this relationship that the prime numbers construct the
$\zeta(k)$ function. Since $p$ denotes a prime number and $k>1$, none of the
factors in this product can be zero. Therefore we can conclude that $\zeta(k)$
does not have any zeros if $k>1$.

\begin{figure}
   \caption{\label{fig:RZOnComplexPlane}%
            The `anatomy' of the Riemann-zeta function on the complex $s$ plane.
            The black dots ($\bullet$) represent the zeros of $\zeta(s)$,
            including possible zeros which do not lie on the critical line.}
   \includegraphics[width=0.45\textwidth]{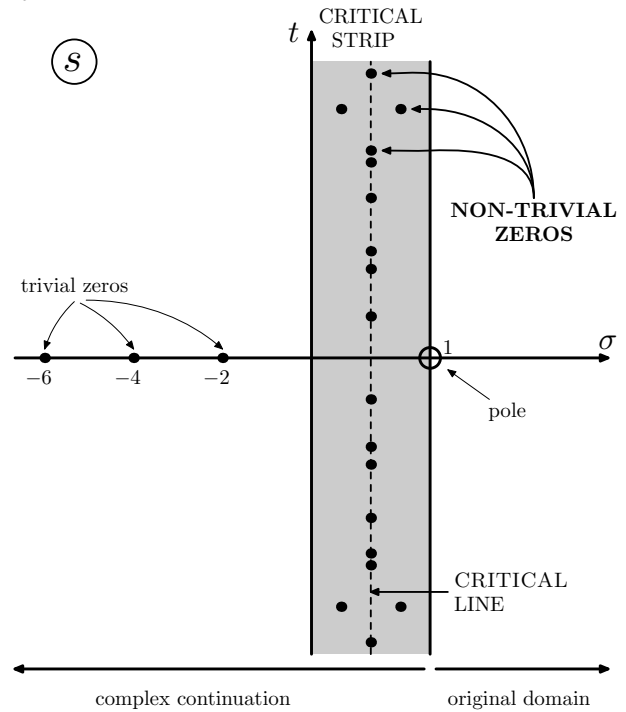}
\end{figure}

Bernhard Riemann, who was the first to apply the tools of complex
analysis to this function, proved that the function defined by the
infinite summation \cite{Riemann1859}
\begin{equation}
   \label{eq:RZDefintion}
   \zeta (s) = \sum_{n=1}^{\infty}{\frac{1}{n^{s}}},
\end{equation}
can be analytically continued over the complex $s$ plane, except for $s=1$.
This analytic continuation of the function is called the Riemann-zeta function.
Here we follow the traditional notation, with $s$ denoting a complex number,
$s=\sigma+it$, where $\sigma$ and $t$ are real numbers and $i$ is the usual
imaginary unit.

Riemann also derived a functional equation, containing the $\zeta(s)$ function,
which is valid for all complex $s$ and exhibits mirror symmetry around the
$\sigma= 1/2$ vertical line, called \emph{the critical line}, such
that
\begin{equation}
   \label{eq:RZFunctionalEquation}
   \pi^{-\frac{s}{2}}
   \Gamma \!\left ( \frac{s}{2} \right )
   \zeta (s) 
   =
   \pi^{-\frac{1-s}{2}}
   \Gamma \!\left ( \frac{1-s}{2} \right )
   \zeta (1-s) 
\end{equation}
One should note that the zeta function stands on both sides, on the
left hand side with argument $s$, while on the right hand side with
($1-s$). This relationship between $\zeta(s)$ and $\zeta(1-s)$ provides
some insight regarding the location of the zeros of this function. Let
us examine the half-line for which $\sigma < 0$, and $t=0$.
The products on either side can be zero if at least one of the factors
is zero. On the right hand side of (\ref{eq:RZFunctionalEquation}) all
the pre-factors of the zeta function are non-negative and do not have
any zeros. On the other side, however, the $\Gamma(\sigma/2)$ function
has simple poles at all even negative integers. The equation can hold
only if $\zeta(\sigma)$ has simple zeros at the same locations. These
zeros are called {\emph{trivial}}, because their locations are inherited
from the $\Gamma$ function. The same argument also shows that all other
zeros of the $\zeta(s)$ function have to lie in the $0 \le \sigma \le 1$
region, called the {\emph{critical strip}}. The zeros located in this
strip are the {\emph{non-trivial}} zeros of the Riemann-zeta function.
It can also be shown that the non-trivial zeros $\rho$ are arranged
symmetrically, both in respect of the critical line and the $t=0$ axis.
Figure~\ref{fig:RZOnComplexPlane} depicts the pole and zero
structure of $\zeta(s)$ on the complex $s$ plane including the possible
zeros off the critical line.

So far the statements about the zeros of $\zeta(s)$ and their
locations on the complex plain were simple. However the distribution
of the non-trivial zeros holds one of the most intriguing and enigmatic
mathematical mysteries of the last century and a half. It is
embarrassingly easy to pose Riemann's conjecture: {\emph{all non-trivial
zeros of $\zeta(s)$ have the form $\rho=1/2+it$, where $t$ is a real
number}}. In other words all non-trivial zeros lie on the critical
line. In 1900 Hilbert nominated the Riemann Hypothesis as the eighth
problem on his famous list of compelling problems in mathematics
\cite{Hilbert1900}. Since then not just professional mathematicians but mathematical
soldiers of fortune tried, and still try, to verify its validity. The stakes
are high. Whoever proves or disproves this hypothesis engraves his name in
the tablets of the history of mathematics, and may also receive one million
dollars from the Clay Mathematics Institute\footnote{See
http://www.claymath.org/millennium/Riemann\_Hypothesis}.

During the past century, the Riemann Hypothesis has been recast
into many equivalent mathematical statements. A few of them are
purely number theoretical in origin, such as the Mertens conjecture,
which we will later discuss in the context of a special Brownian
motion, but other redefinitions are very much cross-disciplinary.
A more advanced mathematical introduction to the history of the
Riemann Hypothesis and its equivalent statements can be found in
an excellent monograph and compendium~\cite{Borwein2008} which is
readable not just at the expert, but also the undergraduate level.

The distribution of the $\zeta(s)$ zeros, with real part equal
to 1/2, has thus attracted significant interest. One of
mathematics' giants has proven that infinitely many zeros do lie
on the critical line~\cite{Hardy1914}, however Riemann's conjecture
is much stronger, requiring all the zeros to be on the critical
line. In 1942 Selberg proved
\begin{equation}
   N_{0}(T) > C\ T \ln{\!(T)}
   \qquad
   (C >0 {\mbox{ and }} T \ge T_{0})
\end{equation}
i.e. the number of zeros of the form $s=\frac{1}{2} + it$
($0 \le t \le T$), denoted by $N_{0}(T)$, grows as $T \ln{(T)}$
at least for large $T$. Three decades later, in 1974, Levinson
showed that at least one third of the non-trivial zeros are on
the critical line~\cite{Levinson1974} which was later incrementally
improved to two fifths~\cite{Conrey1989}. This small step over
a period of twenty years is indicative of the difficulty of the
Riemann Hypothesis.

Let us return to the linkage between the $\zeta(s)$ zeros and prime
numbers. Equation (\ref{eq:EulerProduct}) clearly shows the strong
connection between the $\zeta(s)$ function and the prime numbers.
This relationship can be made even more explicit if one examines how
the number of primes below a given threshold behaves as this threshold
is increased. Based on empirical evidence, many mathematicians, e.g.
Legendre, Gauss, Chebyshev~\cite{Dickson2005}, have conjectured that
the prime counting function, $\pi(x)=\vert \lbrace p \,\vert\, p
\hspace*{1mm} {\mbox{is prime and}} \hspace*{1mm} p \le x \rbrace \vert$,
asymptotically behaves as the logarithmic integral $\mathrm{Li}(x)$.
This conjecture is known nowadays as the Prime Number Theorem after
Hadamard \cite{Hadamard1896} and de la Vall{\'e}e-Poussin~\cite{Poussin1896}
independently gave rigorous proofs of this statement. Interestingly,
this theorem has a geometrical interpretation: the Prime Number Theorem
is equivalent to the assertion that no zeros of $\zeta(s)$ lie on the
$\sigma=1$ boundary of the critical strip.

Riemann published \cite{Riemann1859}, although Mangoldt provided the rigorous
proof~\cite{Mangoldt1895}, the following explicit formula for the
prime-counting function $\pi(x)$
\begin{equation}
   \label{eq:ExactFormulaForPi}
   \pi(x) = \sum_{n=1}^{\infty}{\frac{\mu(n)}{n} J\!\left ( x^{1/n} \right ) }
\end{equation}
where
\begin{eqnarray*}
   J(x)
   =&&
       \mathrm{Li}(x)
     - \lim_{T \rightarrow \infty}{\left \lbrack
                                   \sum_{\vert \rho \vert \le T}{\mathrm{Ei}(\rho \log{(x)})} 
                                   \right \rbrack} + \\
   && + \int_{x}^{\infty}{\frac{dt}{(t^{2}-1)t \log{(t)}}} 
      - \log{(2)}.
\end{eqnarray*}
Here $\mu(n)$ is the M{\"{o}}bius function\footnote{The M{\"o}bius function
is defined as follows: $\mu(1)=1$, $\mu(n)=0$ if $n$ has a square divisor,
and $\mu(p_{1}p_{2}\cdots p_{k})=(-1)^{k}$ if all $p_{i}$s are different.
Thus $\mu(2) =-1$ and $\mu(12)=0$, and $\mu(21)=1$.}, $\rho$ denotes the
non-trivial zeros of the Riemann $\zeta(s)$ function, and $\mathrm{Li}(x)$
and $\mathrm{Ei}(x)$ stand for the logarithmic and exponential integrals%
\footnote{The notation for the logarithmic integral is ambiguous in the
literature. There are two definitions
\begin{equation*}
   I_{1}(x) = \dashint_{0}^{x}{\frac{dt}{\ln(x)}}
   \quad \mbox{and} \quad
   I_{2}(x) = \int_{2}^{x}{\frac{dt}{\ln(x)}}
\end{equation*}
where $I_{1}$ is interpreted as a Cauchy principal value. These integrals
differ only by a constant number. Depending on the book the Reader may
consult, either $I_{1}(x)$ or $I_{2}(x)$ is denoted with $\mathrm{Li}(x)$.
Here, we prefer the former.
}%
, respectively. Therefore, whoever knows the distribution of the non-trivial
zeros of $\zeta(s)$, will also know the distribution of the prime numbers.

\begin{figure}[htb!]
   \includegraphics[angle=-90,width=0.47\textwidth]{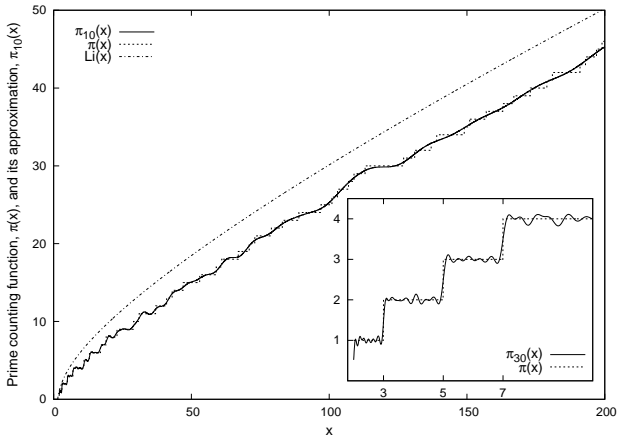}
   \caption{\label{fig:RZ_ExactFormulaForPi}
            Figure depicts the approximation (\ref{eq:ExactFormulaForPi}) for
            the prime counting function, $\pi(x)$ (dashed line), using only the
            first term, ${\mathrm{Li}}(x)$ (dash-dotted line), and using the
            first ten non-trivial pairs of zeros of the Riemann $\zeta(s)$
            function (solid line). In the inset we restricted the range to $[2,
            10]$ and used the first 30 non-trivial zeros.}
\end{figure}

Selecting only the first terms of the summand in equation
(\ref{eq:ExactFormulaForPi}) reproduces exactly the Prime Number
Theorem, i.e.
\begin{equation}
   \pi(x) \cong \mathrm{Li}(x) \cong \frac{x}{\ln{(x)}}\,.
\end{equation}
This observation may lead us to conclude that ${\mathrm{Li}}(x)$ gives the main
contribution to $\pi(x)$ while the other terms represent corrections, similar
to a perturbative calculation in physics -- an analogy to which we will return. 
Figure \ref{fig:RZ_ExactFormulaForPi} depicts the prime counting function,
$\pi(x)$ and its various approximations. One may notice that the leading term,
${\mathrm{Li}}(x)$, captures the tendency of $\pi(x)$ well and the appearance of
the oscillations clearly show how the zeros $\rho_{n}$ influence and refine the
agreement. As $x \rightarrow \infty$ the curves of ${\mathrm{Li}}(x)$ and
$\pi(x)$ will practically coincide on a similar plot.

One may define a density for the complex, non-trivial Riemann-zeta zeros as
\begin{equation}
   d(N) = \sum_{k}{\delta \!\left ( N - \rho_{k} \right )}
\end{equation}
where $\delta$ is the Dirac-delta distribution. Following Sir Michael Berry
\cite{Berry1985} the spectral density can be separated into a smooth and an
oscillatory part, $d(N) = {\bar{d}}(T) + d_{\mathrm{osc}}(T)$, as
\begin{subequations}
\begin{eqnarray}
   \label{eq:RZ_DensityOfZerosSmoothPart}
   {\bar{d}}(T)
   &=& 
   \frac{1}{2\pi} 
   \ln{\!\left ( \frac{T}{2\pi} \right )}
   + 1 - \frac{1}{2\pi} + {\cal{O}}\!\left ( T^{-1} \right )
   \\
   \label{eq:RZ_DensityOfZerosOscillatoryPart}
   d_{\mathrm{osc}}(T)
   &=&
   - \frac{1}{\pi}
   \sum_{p}%
       {
        \sum_{r=1}^{\infty}%
            {
             \frac{
                   \ln{(p)}
                   \cos{ \left ( 
                                 r T \ln{(p)}
                         \right )}
                  }%
                  {
                   \sqrt{p^{r}}
                  }
            } 
       }
\end{eqnarray}
\end{subequations}
where the external summation of $d_{\mathrm{osc}}(T)$ runs over the prime
numbers, $p$. The oscillatory part, therefore, gives the fluctuations as
individual contributions from each prime number $p$ labeled by an integer
$r$ corresponding to the prime power $p^{r}$. Based on the smooth density
of Riemann-zeros one may derive the number of positive, non-trivial zeros
upto a fixed value of $T_{0}$:
\begin{equation} \label{eq:RZCountingFunctionSmoothPart}
   N(\rho < T_{0})
   =
   \int_{0}^{T_{0}}%
       {\bar{d}(T) \, dT}
   =
   \frac{T_{0}}{2\pi}
   \ln{\!\left ( \frac{T_{0}}{2\pi} \right )}
   -
   \frac{T_{0}}{2\pi}
\end{equation}
Changing variable to ${\cal{T}}=\ln{\!\left ( T_{0}/2\pi \right)}$ and
recasting our result using ${\cal{T}}$ we obtain
\begin{equation}
   N({\cal{T}}) \propto e^{\cal{T}}
\end{equation}
i.e. the number of $\zeta(s)$ zeros below ${\cal{T}}$ increases
exponentially. Although at this point this change of variable seems
somewhat arbitrary, we will see later that it further strengthens
the similarity between the zeros of $\zeta(s)$ and the periodic orbits
of a chaotic system, where the number of periodic orbits also increases
exponentially.

Finally, we note the fruitful and diverse area of extensions of
the Riemann-zeta function. These generalised zeta-functions do
also occur throughout physics, primarily in modern quantum field
theories. This topic, however, is far beyond the scope of this
short review and we can only suggest Elizalde's monograph 
\cite{Elizalde1995} as an introduction and Lapidus' book
\cite{Lapidus2008} for a more authoritative study.

\section{\label{sec:ConnectionsToPhysics}Connections to physics}

\hfill \parbox{75mm}{\raggedleft
                     {\emph{The Riemann Hypothesis is a precise statement, and
                            in one sense what it means is clear, but what it's
                            connected with, what it implies, where it comes
                            from, can be very unobvious.}} \\
                            (Martin Huxley)
                     } \\

\subsection{\label{sec:ClassicalMechanics}Classical mechanics}

\noindent {\emph{In this section we discuss those models of classical mechanics, such as
billiards, which lead to the introduction of the notion of integrability and
chaos. This development of ideas gave birth to a new paradigm, since it provided
an insight into how the spectrum of quantised analogues of classical systems
are connected to classical paths.}}

Classical mechanics, in its Lagrangian and Hamiltonian forms, is the exemplar
for physics in the modern sense. The major
theories, e.g. statistical mechanics, quantum mechanics, are first expressed in
the language of analytical mechanics with the development traced to the
Enlightment. Although a few analytically solvable models, e.g. Kepler two-body
problem, harmonic oscillator, gave confidence in the machinery of mechanics, it
was soon realised that there are important cases, e.g. three-body problem, where
one not just cannot solve the equations of motion analytically, but the motion
is proven to be chaotic~\cite{Celletti2007}. This behaviour is very peculiar and
at first sight seems puzzling, since the governing equations are deterministic,
yet the actual motion seems to behave randomly. The celestial relevance of this
three-body problem was so fundamental and enticing that King Oscar II of Sweden
and Norway offered a prize for the person who could solve the following problem
\cite{Barrow-Green1994}
\begin{quotation}
	\noindent For an arbitrary system of mass points which attract
	each other according to Newton's law, assuming that no two points
	ever collide, give the co-ordinates of the individual points for
	all time as a sum of a uniformly convergent series whose terms
	are made up of known functions.
\end{quotation}
Although this problem had not been solved, Poincar{\'e} was awarded
this illustrious prize for his impressive contribution. His work
revolutionised the analysis of such chaotically behaving systems,
although one had to wait nearly a hundred years for this revolution
to really happen.

In classical mechanics we distinguish a special class of systems, the
{\emph{integrable dynamical system}}, which possess as many independent
integrals of motion, ${\cal{I}}_{n}$, (action variables) as degrees of freedom,
$N$. For these systems the Hamiltonian can be expressed as a function of these
action variables, namely ${\cal{H}} = {\cal{H}}({\cal{I}}_{1}, \dots,
{\cal{I}}_{N})$, and the equations of motion ($n=0$, $1$, \dots, $N$)
\begin{equation}
   \frac{d\varphi_{n}}{dt} = -\frac{\partial {\cal{H}}}{\partial I_{n}}
   \quad \mbox{and} \quad
   \frac{d I_{n}}{dt} = \frac{\partial {\cal{H}}}{\partial \varphi_{n}}
\end{equation}
are easy to solve: $I_{n}=$ constant and $\varphi_{n}=\varphi_{n,0} + \omega_{n}
t$. A theorem of topology then guarantees that these $N$ constants of motion,
provided they are independent of each other, define an $N$ dimensional torus
and each trajectory with constant energy lies on that torus. Therefore, as
a specific case, the dynamics described by a one-dimensional time-independent
Hamiltonian is necessarily integrable. In order to consider chaotic dynamics
one has to either introduce a time-dependent Hamiltonian or increase the
degrees of freedom to two or higher.

One of the `simplest' generic models with two or more degrees of freedom is that
of classical billiards. These are dynamical systems where a particle has
constant energy and moves in a finite volume, which may contain impenetrable
obstacles. Whenever the particle reaches the boundary it suffers specular
reflection. Depending on the shape of the billiard, the motion can be integrable
or chaotic.
\begin{figure}[t!]
   \includegraphics[width=0.46\textwidth]{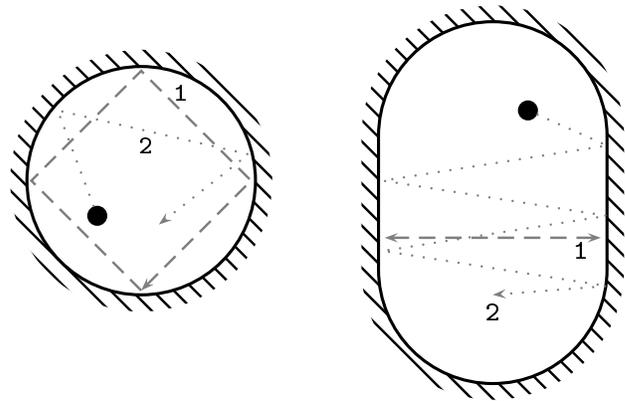}
   \caption{\label{fig:CM_Billiards}
            A circular billiard and a Bunimovich stadium, which is a
            rectangle smoothly joined by semi-circles. Two different
            types of trajectories, periodic orbits (1) and non-periodic
            trajectories (2) are also depicted.}
\end{figure}
The analysis of a circular billiard (see Figure~\ref{fig:CM_Billiards})
is straightforward due to the rotational symmetry. The incident angle remains
the same at each bounce and each impact can be calculated from the
previous one by rotating the circle twice that angle. Therefore if the
incident angle is a rational multiple of $\pi$, i.e. $m\pi/n$, the trajectory
is periodic with period $n$ and therefore finite, otherwise it is infinite.
In this latter case the points where the ball hits the wall will be uniformly
distributed along the circumference of the circle. It was also proven by Jacobi
that in the latter case every interval of the circle contains points of the
trajectory.

\begin{figure}[bht!]
   \includegraphics[width=0.40\textwidth]{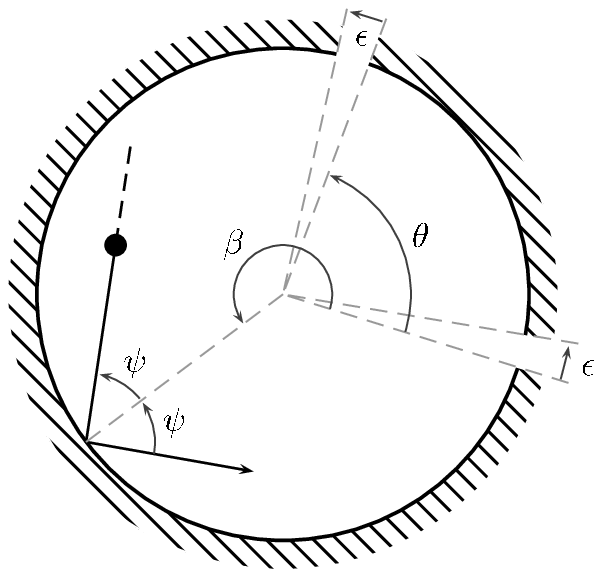}
    \caption{\label{fig:CM_OpenCircularBilliard}
            A circular billiard with to small openings 
            \cite{Bunimovich2005a}.}
\end{figure}

Before we step beyond billiards and generalise the idea of periodic orbits, the
origin of trace formulae, let us make a short detour around a recent result
\cite{Bunimovich2005a} regarding the circular billiard (see Figure
\ref{fig:CM_OpenCircularBilliard}). As we discussed, due to rotational symmetry,
or in other words, the conservation of angular momentum, this billiard model is
integrable and the trajectory is fully described by two angles, $\beta$ and $\psi$,
the angle around the circumference measured from a pre-determined point and the
incident angle of the trajectory at the boundary, respectively. With these
variables the dynamics is governed by the mapping: $(\beta, \psi) \mapsto (\beta
+ \pi - 2\psi, \psi)$, where all angles are taken modulo $2\pi$ and the ball
travels with unit velocity. The phase space of this system can be
described by Birkhoff's coordinates constructed from two angles; the
arc-length coordinate $q=\beta$ (measured in radians and modulo $2\pi$),
and the tangential momentum coordinate defined as $p = \sin{\!(\psi)}$. By convenient
normalisation, the arc-length of the billiard is unity and the velocity of the ball
is also unity, the phase space is restricted to $0 \le q < 2\pi$, and $-1 < p < 1$. This
choice also introduces a natural unit time-step, the time elapsed between consecutive
bounces, $\Delta t = 2 \cos{(\psi)}$. The movement of the ball can, therefore, be
represented by a possibly infinite series of points inside this phase-space area.
Despite the rather artificial appearance of this model, the electromagnetic field
in optical or microwave cavities can be modeled by such billiards \cite{Stockmann1990,
Nockel1997, Alt1998, Harayama2003}. Since these experimental billiards are not ideal,
it is interesting to examine what happens to the dynamics of this system if we cut
a small window(s) along the reflective boundary, thereby, naturally introducing
dissipation or `leakage'. It is natural to ask: what is the probability,
${\mathbf{P}}(n)$, of a ball leaving the billiard after $n$ bounces, what is the
mean number of bounces, $\mean{n}$, before the ball escapes, or similarly,
what is the probability, ${\mathbf{P}}(t)$, that escape takes at least time $t$.

For strongly chaotic billiards the latter probability decays exponentially, while
for integrable billiards, such as the circular one, it softens to only power-law
decay \cite{Bauer1990} and can be qualitatively understood using a simple geometrical
argument. The probability, $p$, that the ball escapes in a bounce is proportional
to the size of the gap to that of the boundary, $p = \epsilon/L$. Moreover, the
probability that the ball survives the first $(n-1)$ bounces and escapes only
at the $n$th bounce is $(1-p)^{(n-1)}p$. Therefore the mean number of bounces
occurring until escape is
\begin{equation}
   \mean{n_{\mathrm{escape}}}
   =
   \sum_{k=1}^{\infty}{k \, (1-p)^{(k-1)}p}
   =
   \frac{1}{p}
   \propto
   \frac{1}{\epsilon}.
\end{equation}

Let us now cut two (possibly overlapping) holes, with sizes $\epsilon$, on the
boundary and examine the non-escaping periodic orbits. Based on the geometrical
argument used above, we expect the probability to be $\sim 2/\epsilon$, if the
two holes do not overlap. However, in systems where the trajectories do not diverge
strongly, i.e. Lyapunov exponent is close to zero, only a small fraction of the
trajectories will eventually hit the opening on the boundary, and the mean escape
time will be proportional to $\epsilon$.

If the initial incident angle is taken to be $\psi_{m,n} = \pi/2 - m\pi/n$, where
$m<n$ are integers and relative primes to each other, then the trajectory is closed
and its period is $n$. Let us now examine only those initial conditions for which
the escape time is at least $t$, or in other words, the number of bounces is at
least $N=\lfloor 2\pi/\epsilon \rfloor$. To fulfill this requirement one might take
the initial value of $\psi = \psi_{m,n} + \eta$, where $0 \le \eta \ll \epsilon$ and
$\beta$ can be restricted to the following range
\begin{equation*}
   \beta_{0}'
   \in
   \left ( 
            \epsilon +
            \frac{\eta t}{\cos{(\psi_{m, n})}} ; \theta'
   \right )
   \! \bigcup \!
   \left (
           \theta'  +
           \epsilon +
           \frac{\eta t}{\cos{(\psi_{m, n})}}; \frac{2\pi}{n}
   \right ).
\end{equation*}
The prime indicates that angles are taken modulo $2\pi /n$. The probability
can, therefore, be calculated if one sums up all possible values of ($m$,
$n$) pairs. This is the point where number theory enters into this physical
problem; we have to guarantee that $m$ and $n$ are relative primes.
Integrating over the permitted region of $\beta'$ one may find
\begin{equation}
   \label{eq:BilliardProbability01}
   {\mathbf{P}}(t, \epsilon, \theta)
   \sim
   \frac{1}{t}
   \sum_{n=1}^{N}%
       {
        n \,\mathcal{F}(n)
        \sum_{m}{\left \lbrack 1 -  \cos{\!\left ( \frac{2m\pi}{n} \right )} \right \rbrack }
       }
\end{equation}
where the exact form of $\mathcal{F}(n)$ can be found in \cite{Bunimovich2005a}.
Surprisingly the sum over $m$ can be explicitly determined. The first, unit term,
simply counts how many numbers are relative prime to $n$ and, therefore, it 
can be formally expressed using a special function of number theory; Euler's
totient function~\footnote{Euler's totient function, $\phi(n)$, gives the number
of positive integers smaller than $n$, which are relative prime to $n$, e.g. for
any prime number $\phi(p)=p-1$, since all integers smaller than $p$ are relative
prime to $p$.}. The second term in equation (\ref{eq:BilliardProbability01}) is
also a special expression. If the summation were over all the integer numbers
smaller than $n$, one could connect it to the Fourier series. However, here
one only uses those $m$'s which are relative primes to $n$. Converting the cosine
term to complex exponentials and using Ramanujan's identity\footnote{Ramanujan's
sum is defined as
\begin{equation*}
   c_{n}(m) = \sum_{m}{e^{2\pi i m/n}}
\end{equation*}
where the summation is over those values of $m$, which are relative prime to $n$.
Using M{\"o}bius inversion for this sum one can prove that $c_{n}(m) = \mu(n)$
\cite{Hardy1960}.} for the sum of exponentials, the contribution of the cosine
term turns out to be another special function of number theory which we have
already met, the M{\"o}bius function, $\mu(n)$. Therefore, the probability of
non-escaping orbits is
\begin{equation}
   {\mathbf{P}}_{\infty}
   =
   \lim_{t \rightarrow \infty}{\!\bigl ( t {\mathbf{P}}(t, \epsilon, \theta) \bigr )}
   \sim
   \sum_{n=1}^{\infty}%
       {
        n \left \lbrack \phi(n) - \mu(n) \right \rbrack {\mathcal{F}}(n)
       }
\end{equation}
The leading order behaviour of $P_{\infty}$ as a function of $\epsilon$ can be
determined by calculating its Mellin-transform
\begin{equation}
   \widetilde{{\mathbf{P}}}(s)
   =
   \int_{0}^{\infty}{{\mathbf{P}}_{\infty}(\epsilon, \theta) \, \epsilon^{s-1}\, d\epsilon}
\end{equation}
and examining the residues of $\widetilde{P}(s)$ on the complex $s$-plane. Bunimovich
and Detteman showed that for the two-hole problem, where these holes are separated
by $0^{\circ}$, $60^{\circ}$, $90^{\circ}$, $120^{\circ}$, $180^{\circ}$, the probability
$\widetilde{{\mathbf{P}}}(s)$ is uniquely determined by the Riemann-zeta function, $\zeta(s)$,
by i.e. its pole and non-trivial zeros. The first corrections to the leading order
term are given by the non-trivial zeros of $\zeta(1+s)$, which are of the order
$\sqrt{\epsilon} \, \ln{\!(\epsilon)}^{m-1}$, provided for all zeros of $\zeta(s)$,
$\Re{(s)} = \sigma \le \frac{1}{2}$ with multiplicity $m$. The Riemann Hypothesis
is then shown to be equivalent to different asymptotic estimates on the number of zeros
\cite{Titchmarsh2003}. Therefore, if the non-trivial zeros provide the second order
corrections to the probability, it is instructive to examine the deviation of
these probabilities experimentally from the leading-order geometric terms, namely
\begin{subequations}
\begin{eqnarray}
   \lim_{\epsilon \rightarrow 0}%
       {
        \lim_{t \rightarrow \infty}%
            {
             \left (
                    \epsilon^{\delta - 1/2}
                    \left \lbrack 
                        t {\mathbf{P}}_{1}(t) - \frac{2}{\epsilon}
                    \right \rbrack
             \right )
            }
       }
   = 0
   \\
   \lim_{\epsilon \rightarrow 0}%
       {
        \lim_{t \rightarrow \infty}%
            {
             \left (
                    \epsilon^{\delta - 1/2}
                    \biggl \lbrack 
                        t {\mathbf{P}}_{1}(t) - 2t {\mathbf{P}}_{2}(t)
                    \biggr \rbrack
             \right )
            }
       }
   = 0
\end{eqnarray}
\end{subequations}
where ${\mathbf{P}}_{1}$, ${\mathbf{P}}_{2}$ belong to the one- and two-hole
problem, respectively. If it is (experimentally) found that for every $\delta
> 0$ these equations are fulfilled, then it proves the validity of the asymptotic
formulae, thus the validity of the Riemann Hypothesis. The numerical results
by Bunimovich and Detteman does not contradict these equations. Although their
result has not proven the Riemann Hypothesis, it provides a physically realisable
system where actual measurements can substantiate, but not prove, the conjecture.

Let us turn our attention now to the dynamics of a more general billiard system. If
one smoothly deforms the boundary of this circle and creates a stadium-like shape,
the analysis is far less straightforward~\cite{Bunimovich1979}. However, qualitatively
we may see that
the trajectories can be classified similarly and one may distinguish periodic
orbits, i.e. $\lbrace q_{n}(t_{1}), p_{n}(t_{1}) \rbrace = \lbrace q_{n}(t_{2}),
p_{n}(t_{2}) \rbrace$ for $t_{1} < t_{2}$, and non-periodic trajectories. It is
tempting to think that periodic orbits are exceptional and quite rare among all
orbits, since for integrable systems the number of periodic orbits grows
polynomially and one may expect that violating integrability would decrease the
number of periodic orbits. In fact, the opposite is true. These special orbits
proliferate among the possible orbits and their number, for a general
Hamiltonian dynamics, grows exponentially with the length of the periodic
orbits, $\sim e^{h \ell}/h \ell$, where $h$ is called the topological
entropy and $\ell$ denotes the length of a given class of periodic
orbits~\cite{GutzwillerBook, Stockmann1999}. This is a striking difference
between integrable and chaotic systems. It is even more surprising that the knowledge
of these periodic orbits serves as a powerful analytical tool for investigating
chaotic systems, and, moreover, they provide the pathway, through trace formulae,
from classical to quantum mechanics.

Let us examine the time evolution of a Hamiltonian flow in general. We denote
the trajectory starting its time evolution from the initial point
${\mathbf{r}}_{0}$ with ${\mathbf{r}}(t) = {\mathbf{F}}({\mathbf{r}}_{0}, t)$.
We further introduce the {\emph{evolution operator}} with the following
definition
\begin{equation}
   {\cal{L}}(t; {\mathbf{r}}', {\mathbf{r}})
   =
   \delta
      \!\left (
         {\mathbf{r}}' - {\mathbf{r}}(t)
      \right )
   =
   \delta
      \!\left (
         {\mathbf{r}}' - {\mathbf{F}}({\mathbf{r}}, t)
      \right ).
\end{equation}
It can be shown rigorously for a generic classical chaotic system
\cite{Cvitanovic1991}, that
\begin{equation}
   \tr{\left ( {\cal{L}}(t; {\mathbf{r}}', {\mathbf{r}}) \right )}
   =
   \sum_{p}%
   {T_{p}
    \sum_{r=1}^{\infty}%
    {\frac{\delta{\left ( t - r T_{p} \right )}}%
          {\abs{\det{\left ( {\mathbf{1}} - {\mathbf{J}}_{p}^{r} \right )}}}
   }}.
\end{equation}
where the first summation runs over the periodic orbits labeled by
$p$, while the second takes into account all repetitions, $r$.
${\mathbf{J}}_{p}$ is the Jacobian matrix of ${\mathbf{F}}$ localised
around the periodic orbit, also called the {\emph{monodromy matrix}}.

Here we can make an important observation: although this equation looks
cumbersome, it does relate the spectrum of the evolution operator to a
global behaviour of periodic orbits. Therefore these two sets of abstract
objects are intimately related to one another. The connection
of this trace formula and its quantum mechanical counterpart to the
Riemann zeta function will become clear in the next section.

\subsection{\label{sec:QuantumMechanics}Quantum mechanics}

\noindent
{\emph{Below we expound the P{\'{o}}lya-Hilbert conjecture. We enumerate
the one-dimensional Hamiltonians proposed for which the distribution of
energy eigenvalues mimic the non-trivial zeros of the Riemann zeta function
and analyse their relationship with the Gutzwiller trace formula.
We also examine the possible symmetries of a `Riemann-operator' since it
partially encouraged the development of quantum mechanics with only CT or PT
symmetry.}}

In the dawn of the 20th century Bohr postulated a series of rules for describing
the spectrum of the hydrogen atom well before the birth of Schr{\"o}dinger's and
Heisenberg's quantum mechanics. In these early days `quantisation' meant to
restrict the possible values of action variables of the classical system
(Bohr-Sommerfeld, Wentzel-Kramers-Brillouin, etc.) and the rules worked well, upto an
additive constant. However, this description cannot be satisfactory in general,
since for the majority of classical systems the only constant of motion is the
energy, and therefore a method of quantisation relying on the existence of
action-angle variables could not be applied \cite{Einstein1917b}. On the other hand,
we know that classical mechanics works well for large systems, therefore quantum
mechanics must give the same predictions for a large system as classical mechanics
(Bohr's correspondence principle). This unproven principle ties these two theories
firmly together and the same principle inspired the use of the Riemann $\zeta$
function in investigating the relationship of classical to quantum mechanics.

The basic question is: how can we quantise a classical mechanical system? Could
we state anything about the spectrum of a quantum system, at least qualitatively,
without solving the corresponding Schr{\"o}dinger equation?

We cannot expect to be able to infer the complete spectrum of a generic
system, but asking only for the average density of states may prove
feasible. One can give a crude, although remarkably precise, estimate:
each quantum state occupies approximately $\hbar^{f}$ phase-space, where
$\hbar$ is the Planck constant divided by $2\pi$ and $f$ is the number
of degrees of freedom. This result is rather general and the individual
quantum systems differ only in the `fluctuations' around this average.
It turns out that the type of fluctuation depends on the behaviour of
the classical counterpart; classically regular and chaotic systems are
quite different. For example, a classical rectangular billiard is
integrable (regular), whereas its quantum analogue exhibits a chaotic
spectrum, with the spacing of the quantum levels, $s = \epsilon_{n}
- \epsilon_{n-1}$, following an exponential distribution ${\mathbf{P}}(s)
\sim e^{-s}$ (see later in section \ref{sec:NuclearPhysics}). However,
a stadium billiard is classically chaotic, but the spectrum of its
quantum counterpart is regular, meaning that ${\mathbf{P}}(s)$ is
small for small values of $s$ and sharply peaked at a finite value
indicating a regular distribution of energy levels. We will see that
interpreting the $\zeta(s)$ zeros as energy levels their distribution
is breathtakingly similar to those of a quantum system's. This has
inspired physicists to examine whether one could associate a dynamical
system with the Riemann zeta function.

The advantage of this approach would be that the huge number of $\zeta(s)$
zeros are known and quick numerical algorithms have also been developed to
find further zeros, thus solving the Schr{\"o}dinger equation for large
energies would be unnecessary. The Riemann zeta function could play the
same role in the examination of chaotic quantum systems as the harmonic
oscillator does for integrable quantum systems. This is the point where
the examination of the Riemann zeta function may help to understand physics
or, vice versa, the physics may lead us to the solution of this so far
intractable mathematical problem.

In order to establish a strong formal connection between a generic chaotic quantum
system and the distribution of the Riemann $\zeta(s)$ zeros, we have to elucidate a
new description of quantum systems, the Gutzwiller's trace formula. This trace
formula is the analogue of equations (\ref{eq:RZ_DensityOfZerosSmoothPart}-b) for
physical systems.

Let us, therefore, return to a classically integrable system, for which the
Hamiltonian ${\cal{H}}$ can be given in terms of conserved quantities ${\cal{H}}
= {\cal{H}}({\cal{I}}_{1}, \dots, {\cal{I}}_{N})$. Using Bohr's semiclassical
quantisation rules, these action variables take not arbitrary, but fixed values
\begin{equation}
   {\cal{I}}_{k} = \hbar \left ( n_{k} + \frac{\mu_{k}}{4} \right ),
   \qquad (k=1,2,\dots,N)
\end{equation}
where the $\mu_{k}$ are integers and called Maslov indices~\cite{Arnold1997}.
The density of states, therefore, becomes
\begin{equation}
   d(E)
   =
   \sum_{\mathbf{n}}%
       {\!
        \delta \!\left (
                      E - {\cal{H}} \left ( {\mathbf{I}} \right )
               \right )
        }.
\end{equation}
which can be recast as the sum of a smooth and an oscillatory term.
The former originates from the Thomas-Fermi semi-classical approximation
\begin{equation}
   d_{\mathrm{TF}}(E)
   =
   \int{
        \delta \!\left (
                        E - {\cal{H}} \left ( {\mathbf{p}}, {\mathbf{q}} \right )
                 \right )
        \frac{d{\mathbf{p}} d{\mathbf{q}}}{(2\pi\hbar)^{f}}
       }.
\end{equation}
while the latter is obtained by expanding the effective action
to quadratic order around the classical periodic orbits
\cite{Berry1976, Frontiers2006}:
\begin{eqnarray}
   \hspace*{-5mm}
   d_{\mathrm{osc}}(E)
   =
   \sum_{\mathbf{N}}%
       {
        \left (
               \frac{2\pi}%
                    {\hbar T_{p}}
        \right )^{(f-3)/2}
        \frac{1}%
             {\hbar^2 \sqrt{\det{\left ( {\mathbf{N}}{\mathbf{Q}}_{i,j} {\mathbf{N}} \right ) }}}
        \, \times}
   \nonumber \\
        \times \exp{\!\left (
                i \frac{S_{p}}{\hbar} - i \frac{\pi}{4} {\mathbf{N}} {\boldsymbol{\mu}} + i \frac{\pi}{4} \beta
               \right )},
\end{eqnarray}
where ${\mathbf{Q}}_{i,j} = \det{({\cal{H}})} \times {\cal{H}}_{i,j}^{-1}$ is the
co-matrix of ${\cal{H}}_{i,j} = \partial_{I_{i}} \partial_{I_{j}} {\cal{H}}$,
while $\beta$ is related to the signature of ${\cal{H}}_{i,j}$.

We see, as in classical mechanics, one can also express the density of states as
a sum of a smooth function $d_{\mathrm{TF}}(E)$ and an oscillatory function
which is defined on the periodic orbits of the semiclassical system. Due to the
correspondence principle, we expect the Thomas-Fermi density of states to remain
valid and only the oscillatory part to vary compared to the semi-classical
derivation.

\begin{figure}[htb!]
   \includegraphics[width=0.47\textwidth]{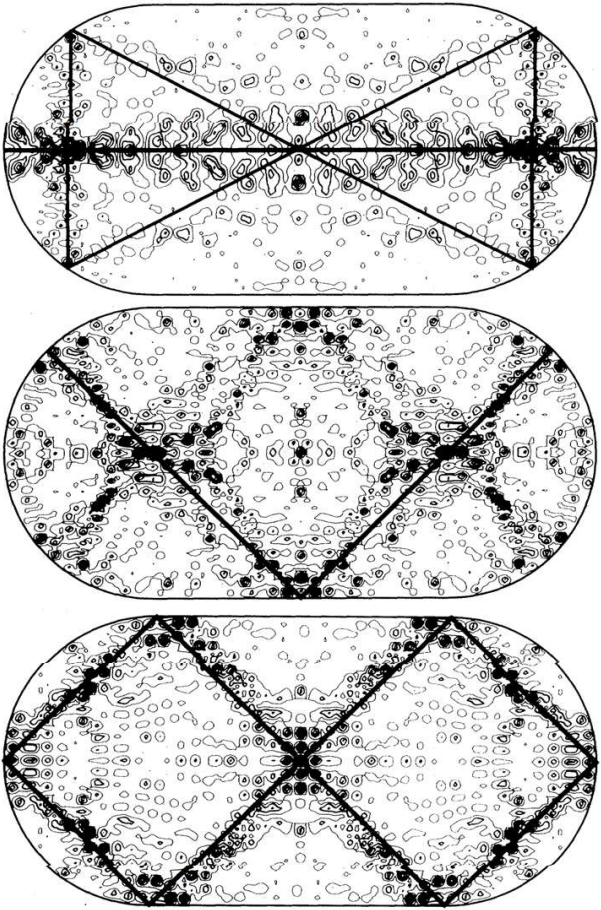}
   \caption{\label{fig:QM_ScarsInStadiumBilliard}
            Three eigenstates of the quantum stadium billiard are shown together
            with the major contributing unstable periodic orbits of the
            classical counterpart as thick solid lines. In the middle figure
            the guiding straight line for the $\wedge$ shaped periodic orbit is
            omitted. From \cite{Heller1984} with the kind permission of the
            author.}
\end{figure}

For non-integrable systems, however, the orbits no longer lie on invariant tori
and a different method is needed for the evaluation of the trace
\begin{equation}
   d(E)
   =
   - \frac{1}{\pi}
   \tr{
        \left ( 
                \Im{
                    \left ( 
                           G_{E}({\mathbf{r}}, {\mathbf{r}})
                    \right )
                   }
        \right )
       }
\end{equation}
where $G_{E}({\mathbf{r}}, {\mathbf{r}})$ is the Green-function associated with
a given Hamiltonian ${\cal{H}}$. This new approach, based on the Green-function,
was developed by \cite{Gutzwiller1970, Gutzwiller1971}. Here
we shall not follow the details of the derivation, but only present the final, fully
quantum mechanical expression for the density of states
\begin{equation} \label{eq:QMGutzwillerTraceFormula_dosc}
   d_{\mathrm{osc}}(E)
   =
   \sum_{\mathrm{p.p.o.}}%
       {
        \frac{T_{p}}{\pi \hbar}
        \sum_{n=1}^{\infty}%
            {
             \frac{\cos{ 
                        \left ( n 
                                \left \lbrack 
                                    \frac{S_{p}}{\hbar} - \frac{\pi}{2} \mu_{p}
                                \right \rbrack 
                        \right )
                       }
                  }%
                  {\abs{
                        \det{
                             \left (
                                 {\mathbf{M}}_{p}^{n}-1 
                             \right )
                            }
                       }^{1/2}
                  }%
            }
       }
\end{equation}
where the summation runs over all primitive periodic orbits, and
${\mathbf{M}}_{p}$ is the monodromy matrix for these primitive periodic orbits.
Using this new method one can derive a semiclassical expression for the spectrum
of a quantum system whose classical analogue is chaotic, when the usual
Bohr-Sommerfeld quantisation rules cannot be applied. Gutzwiller's result above,
therefore, can be viewed as a bridge between the classical and quantum behaviour
of a system, and can provide a rule as to how to quantise such a system. In this
interpretation, Gutzwiller's approach is similar to Feynman's path integral
description, where the quantum system is described in terms of an infinite sum
over classical paths. For the interested reader we can suggest, without any
reservation, Gutzwiller's comprehensive book on classical and quantum
chaos~\cite{GutzwillerBook} and Brack and Bhaduri's monograph giving
an overview of semiclassical physics~\cite{Brack2003}. In order to help the
reader to visualise the emergence of periodic orbits in a quantum mechanical
system we reproduce here a few quantum `scars' from Heller's numerical study.
Figure \ref{fig:QM_ScarsInStadiumBilliard} shows the probability distribution
for three quantum eigenstates of the Bunimovich billiard. It is apparent how the
isolated, unstable classical periodic orbits manifest themselves as paths along
which the probability distribution is greatly enhanced. Gutzwiller's idea to
extract eigenvalues of a chaotic system via the periodic orbits, therefore,
seems most plausible.

Based on analogy between the oscillatory part of the semiclassical density of
states (\ref{eq:QMGutzwillerTraceFormula_dosc}) and the similar expression of
(\ref{eq:RZ_DensityOfZerosOscillatoryPart}) one can set up a
dictionary~\cite{Berry1999a, Bohigas2005} which maps the Riemann zeta function
onto a so-far unknown chaotic quantum mechanical system.
\begin{table}
\caption{Dictionary for translating the `Riemann dynamics' onto a chaotic
         quantum dynamics. Based on~\cite{Berry1999a, Bohigas2005, Brack2003}.
         \label{tab:Dictionary}}
\vspace*{2mm}
\begin{tabular*}{0.47\textwidth}[t]{p{32mm}p{25mm}p{25mm}}
                                & \parbox{25mm}{\centering Generic chaotic system} & \parbox{25mm}{\centering Riemann zeta function} \\[2mm]
   \hline
   periodic orbit labels        & integers               & primes                      \\
   dimensionless action         & $S_{p} / \hbar$        & $T \ln{\!(p)}$              \\
   periods                      & $T_{p}$                & $\ln{(p)}$                  \\
   stability factor$^{\ast}$    & $\det{\left (
                                  {\mathbf{M}}_{p}^{n}-1 
                                  \right )}$             & $p^{r}$                     \\
   Maslov index$^{\ast}$        & $\mu_{p}$              & 2$^{\dag}$                  \\
   asymptotic limit             & $\hbar \rightarrow 0$  & $T_{p} \rightarrow \infty$  \\
   \hline\hline
\end{tabular*}\\[1mm]
\parbox{0.47\textwidth}{\raggedright
                        \footnotesize $^{\ast}$ Depending on how one maps
                        the oscillatory part of the zeta zeros density
                        (\ref{eq:RZ_DensityOfZerosOscillatoryPart}) onto
                        Gutzwiller's trace formula
                        (\ref{eq:QMGutzwillerTraceFormula_dosc}) the definition
                        of the stability factor and the Maslov index can be
                        different. Here we followed \cite{Brack2003}, while
                        another mapping can be found in \cite{Berry1999a}.
             }\\[1mm]
\parbox{0.47\textwidth}{\raggedright
                        \footnotesize $^{\dag}$ Therefore the Maslov phase is
                        $\pi$, but this is not unique and one could also choose
                        $3\pi$, $5\pi$, etc.
             }
\end{table}

Although the Hamiltonian, ${\cal{H}}$, which would describe the chaotic quantum
system corresponding to the Riemann zeta function, is still missing, the mutual
resemblance of (\ref{eq:QMGutzwillerTraceFormula_dosc}) and
(\ref{eq:RZ_DensityOfZerosOscillatoryPart}) reveals some possible properties of
${\cal{H}}$. In \cite{Berry1999a} a thorough and concise summary of these
properties can be found from which we cite but a few for later use:
\begin{enumerate}
   \item{\label{item:Berry1}
         ${\cal{H}}$ has a classical counterpart, since the absence of any
         analogue of $\hbar$ from (\ref{eq:RZ_DensityOfZerosOscillatoryPart})
         indicates the scaling of the dynamics, namely the trajectories are the
         same at all energy scale.         
        }
   \item{\label{item:Berry2}The Riemann dynamics is chaotic and unstable.}
   \item{\label{item:Berry3}The dynamics lacks time-reversal symmetry.}
   \item{\label{item:Berry4}
         The dynamics is quasi one-dimensional, because for a generic $d$
         dimensional scaling system the number of energy eigenvalues increases
         as $\sim E^{d}$ while for $\zeta(s)$ the number of zeros $T < N(T)
         \sim T \ln{(T)} < T^{2}$. Moreover, the appearance of $\sqrt{p^{r}}$ in
         the denominator implies one expanding direction and no contracting one.
        }
\end{enumerate}
Below we pursue the proposed dynamics related to the Riemann zeta function.

In the early days of quantum mechanics Hilbert and P{\'o}lya suggested
a physical way to verify Riemann's Hypothesis:
\begin{quotation}
   \noindent I spent two years in G{\"o}ttingen ending around the begin of 1914.
   I tried to learn analytic number theory from Landau. He asked me one day:
   ``You know some physics. Do you know a physical reason that the Riemann
   Hypothesis should be true.'' This would be the case, I answered, if the
   nontrivial zeros of the Xi-function\footnote{G. P{\'o}lya refers here to the
   Riemann $\zeta(s)$ function.} were so connected with the physical problem
   that the Riemann Hypothesis would be equivalent to the fact that all the
   eigenvalues of the physical problem are real. 

   I never published this remark, but somehow it became known and it is still
   remembered.
   (Private letter to Odlyzko\footnote{See the scanned pages on Odlyzko's
   personal website: \\ {\texttt{http://www.dtc.umn.edu/$\sim$odlyzko/polya/}}}.)
\end{quotation}
The zeros of $\zeta(s)$ can be the spectrum of an operator, ${\cal{R}} = 
\frac{1}{2}\, {\cal{I}} + {\mathrm{i}} {\cal{H}}$, where ${\cal{H}}$ is
self-adjoint. This operator ${\cal{H}}$ might have an interpretation as a
Hamiltonian of a physical system and, therefore, the key to the proof of the
Riemann Hypothesis may have been coded in physics. Since the first occurrence of
this conjecture a number of models have been promoted. Below we separate the
models depending on whether they relate the zeros to the positive energy
spectrum, i.e. the scattering states of a physical system, or to the negative
energy spectrum, i.e. to the bound states of a quantum system.

\subsubsection{\label{seq:qm_scattering}Scattering state models}

Let us first consider the possibility that the Riemann zeta function
is associated with a quantum scattering problem.

A few decades after Riemann created a new geometry with his revolutionary
work \cite{Riemann1867}, Hadamard examined the geodesics, the trajectories
of freely moving bodies, on surfaces with negative curvature in detail
\cite{Hadamard1898} and noticed the occurrence of families of geodesics whose
cross-section exhibits a fractal-like structure, as we would call it nowadays.
These geodesics diverge exponentially, thus the distance between two trajectories,
$\delta(t)$, however small initially, will grow exponentially, $\delta(t) \approx
e^{\lambda t} \delta(0)$, where $\lambda$ is a positive number, called the
{\emph{Lyapunov exponent}}. This sensitivity of the system to the initial
conditions, however, would not necessarily result in chaotic behaviour,
provided the space for the trajectories is infinite. However, if the surface
is compact, the trajectories cannot escape to infinity, rather mix on this
surface. If one wishes to visualise a particular example, consider a donut
with two holes. On this surface the trajectories remain bounded on the surface
without the length of a geodesics being limited \cite{GutzwillerBook, Balazs1986,
Bogomolny1995a}. These two properties, exponential sensitivity of the initial
conditions and mixing, are the main requirements for chaotic motion \cite{ChaosBook}.
The relative simplicity of the description of such surfaces with negative curvature,
and the presence of completely chaotic classical motion motivated several authors
in the mid-1980s \cite{Gutzwiller1983, Balazs1986, Berry1987} to examine how
such a system can be quantised, i.e. what properties do the solutions and eigenvalues
of the equation ${\mathcal{H}} \phi = \lambda \phi$ possess.

More precisely, for free motion, one seeks the solution of
\begin{equation}
   \label{eq:SchrodingerEquationMathsNotation}
   - \Delta \phi_{n} = \lambda_{n} \phi_{n}
\end{equation}
where $\phi_{n}$ are required to be square integrable and the appropriate
boundary conditions are also provided. Over a compact domain equation
(\ref{eq:SchrodingerEquationMathsNotation}) has only discrete eigenvalues.
On a surface with negative curvature, the non-euclidean Green's theorem
shows that the eigenvalues must have
the form $\lambda_{n} = \frac{1}{2} + i \rho_{n}$ ($\rho$ is real) \cite{Gelfand1959}. This
resemblance immediately suggests a connection with the zeros of the Riemann
$\zeta(s)$. It is also proven that, for a compact surface, the set of $n$'s
is finite, but for a non-compact surface, a continuous part of the spectrum
can also appear. In the latter case the scattering (continuous spectrum) is
non-conventional, because it is the result of the geometry (curvature,
compactedness) and not the physical interaction between particles.

In order to express the eigenvalue density the Green-function
is needed. Interestingly, on a surface with negative curvature
the Green-function can be explicitly written as a sum of individual
Green-functions corresponding to the periodic orbits. It is also
a fact that, all periodic orbits are unstable and their action is
$S(E)=k \ell$, where $k$ is the momentum related to the energy
by $2mE/\hbar^{2} = k^{2} + 1/4$ and $\ell$ defines the length
of a closed geodesic belonging to a given conjugacy class. In
this geometry the density of states is expressed by the Selberg
trace formula \cite{Selberg1949}
\begin{equation}
   \bar{\rho}(k)
   =
   \frac{A}{2\pi} k \tanh{(k\pi)}
   +
   \frac{1}{2\pi}
   \sum_{[p]}
       {
       \sum_{n=1}^{\infty}
           {
            \frac{\ell_{p} \cos{(nk \ell_{p})}}
                 {\sinh{(n \ell_{p}/2)}}
           }
       }
\end{equation}
where $A$ is the area of the surface, the first summation runs over conjugacy
classes of primitive elements $p$, the second, their repetitions. It
is important to note, the Selberg trace formula holds {\emph{exactly}}, in
contrast to other trace formulae, because no semi-classical approximation
has been applied, although its convergence property is similar to the
Gutzwiller form: for large $k$ the Selberg and Gutzwiller trace
formulae converge, since the metric is locally Euclidean and waves with short
wavelength lose their sensitivity to the local curvature of the metric. In this
system, the transient scattering states were examined by Pavlov and Fadeev who
related the nontrivial zeros of the zeta function to the complex poles of the
scattering matrix \cite{Pavlov1975}:
\begin{equation}
   S(k)
   =
   \pi^{-2i k}\
   \frac{\Gamma{\left ( \frac{1}{2} + i k\right )} \zeta(1+2ik)}%
        {\Gamma{\left ( \frac{1}{2} - i k\right )} \zeta(1-2ik)}.
\end{equation}
Despite this natural occurrence of the Riemann zeta function and its non-trivial zeros,
no further insight into the zeros has been gained via this route. Detailed discussion
of the Selberg trace formula can be found in \cite{Hejhal1976, Hejhal1983} or more
physics oriented approaches in \cite{Stockmann1999} and \cite{Wardlaw1989} and in the
context of the Casimir-effect in \cite{Elizalde1993a, Kurokawa2002, Schaden2006}.

So, let us return to the scattering formalism in the standard Euclidean
space. Joffily, motivated by Pavlov and Fadeev~\cite{Pavlov1975}, examined the scattering
states of a non-relativistic, spinless particle under the influence of a spherically
symmetric, local and finite potential. He examined the Jost solutions of this scattering
problem \cite{Joffily2003}, which differ from the physical solution of the Schr{\"o}dinger
equation in their asymptotics\footnote{The Jost functions are the solutions of the
Schr{\"o}dinger equation with the following asymptotic behaviour:
\begin{equation*}
   \lim_{x \rightarrow \infty}{\!\left ( e^{ikx} f(\lambda, k, x)\right )}  = 1
\end{equation*}
where $\lambda= \ell + \frac{1}{2}$ is the shifted angular momentum, $k \sim \sqrt{E}$
and $x \in (-\infty, \infty)$. This choice of the boundary condition is motivated by our
physical picture, i.e. the particle should be represented by free plane waves far from
the local potential. The real physical solution of the Schr{\"o}dinger equation
can be expressed as a linear combination of the two Jost functions.}
\cite{Alfaro1965, Newton1982}. In standard non-relativistic scattering
theory the $S$-matrix is given by
\begin{equation}
   S(k) = e^{2i\delta(k)} = \frac{f_{-}(k)}{f_{+}(k)}
\end{equation}
where $\delta(k)$ is the phase shift, and $f_{\pm}(k)$ are the Jost
solutions defined by their boundary conditions $\lim_{r \rightarrow 
\infty}{(f_{\pm}(k) e^{\mp ikr})} =1$~\cite{Alfaro1965}. Provided the
potential has a finite range and decreases sufficiently rapidly, the
Jost solution $f_{+}(k)$ is proven to have infinitely many zeros,
corresponding to the solutions of the Schr{\"o}dinger equation as
outgoing or incoming waves. Resonances (i.e. states with finite
lifetime) occur if $S(k)$ has poles on the complex $k$ plane with
negative imaginary parts: $k_{n}^{2} = \epsilon_{n} - i \Gamma_{n}/2$,
where $\epsilon_{n}$ and $\Gamma_{n}$ stand for the energy and inverse
lifetime associated with the $n$th state. Joffily introduces a mapping
between these zeros of $f_{+}(k)$ onto the critical line and shows
they coincide with the non-trivial zeros of the Riemann zeta function.
He associates this artificial system with a vacuum and the zeros are
interpreted as an infinity of virtual resonances, and thus reflect the
chaotic nature of the vacuum \cite{Joffily2003, Joffily2004}. This
interpretation has also been extended using relativistic scattering
\cite{Joffily2007}.

In another scattering based approach, Chadan and Musette analysed the so-called
`coupling constant spectrum' of a radially symmetric three-dimensional
Hamiltonian~\cite{Chadan1993} where the potential is chosen from a singular
family of functions
\begin{equation}
   {\cal{H}}_{\mathrm{CM}}
   =
   - \frac{d^{2}}{dr^{2}}
   - \frac{\ell(\ell+1)}{r^{2}}
   + \frac{1}{r^2} f_{\mathrm{CM}}
\end{equation}
where $f_{\mathrm{CM}}$ has logarithmic singularities at $r=0$. They
argued that the coupling constant spectrum coincides `approximately'
with the non-trivial Riemann zeros if the problem is restricted to a
finite, closed interval $r \in [0, e^{-4\pi/3}]$. Mathematically
rigorous detailed analysis and the extension of the potential family
was carried out by \cite{Khuri2002}. Furthermore, the existence of a
three dimensional potential, $U_{R}(r)$, was derived whose $s$-wave
scattering amplitude has the complex zeros of the Riemann zeta function
as ``redundant poles''. Examination of $\zeta(s)$ using a quantum
scattering approach is further motivated if one compares the plot of
the phase of $\zeta(s)$ on the complex plane with with the usual
Argand-diagram\footnote{Argand diagrams can be thought of as a
parametric plot of the inherently complex scattering amplitude
on the complex plane, and the collision energy plays the role
of the parameter (see for example \cite{Bohm2001} or
\cite{Bhaduri1988}).} of the scattering amplitude corresponding
to a collision.
%
\begin{figure}
   \includegraphics[angle=-90, width=0.49\textwidth]{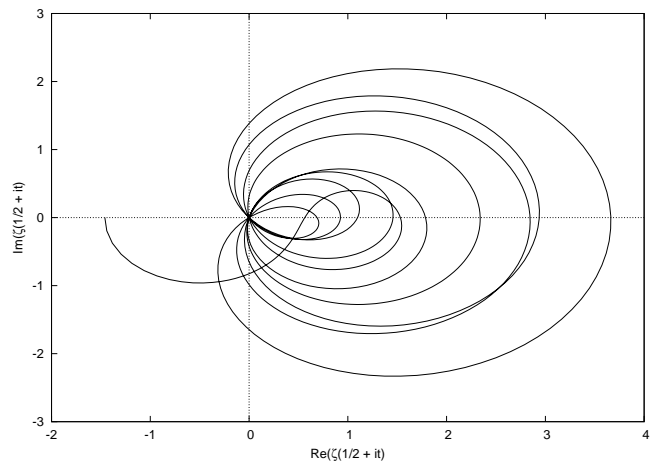}
   \caption{\label{fig:QM_RiemannZetaArgandDiagram}
            Argand diagram of the Riemann zeta function on the critical line,
            $\zeta(1/2 + i t)$, where $t=0-$49.77, the later of which is 
            $\approx \rho_{10}$.}
\end{figure}

As a specific example, for completely elastic collisions, the
scattering amplitude should be a perfect circle on the complex
plane with unit radius centred on (0,1). For inelastic collisions
this circle deforms. The phase of $\zeta(s)$, after interchanging
the roles of the real and imaginary axes, qualitatively resembles
the Argand diagram of a scattering amplitude. This geometric
similarity suggests an analysis of $\zeta(s)$ as if it represented
the scattering amplitude of a real collision of particles. This
analogy, however, is not perfect since $\zeta(s)$ does become negative
while the Argand diagram of the scattering amplitude corresponding to
a realistic collision does not. Bhaduri advocates neglecting these
small differences which do not affect their most important result,
namely the phase $\theta(t)$ of the Riemann zeta function along the
critical line, $\zeta(1/2+it) = Z(t) e^{-i \theta(t)}$, is intimately
connected to the quantum scattering of a particle on a saddle-like
surface~\cite{Bhaduri1995, Bhaduri1997a}.

To illustrate this, let us consider a non-relativistic particle moving in an
inverted harmonic oscillator potential along the half-line ($x \ge 0$). The
Schr{\"o}dinger equation reads as
\begin{equation}
   -\frac{\hbar^{2}}{2m} \frac{d^{2}}{dx^{2}} \Phi(x)
   -\frac{1}{2} m \omega^{2} x^{2} \Phi(x) = E \Phi(x)
\end{equation}
where we require that $\Phi(x=0)=0$. This problem can be mapped onto a repulsive
Coulomb problem of which the phase shift $\delta(t)$ can be exactly expressed
\cite{Flugge1974}. The oscillatory part of the phase shift is given by
\begin{eqnarray}
   \delta(t)
   =
   \delta_{\mathrm{smooth}}(t)
   + \hspace*{45mm}
   \nonumber \\
   \Im{
       \!\left \lbrack
          \ln{\!\left ( \Gamma{\left ( \frac{1}{4} + i \frac{t}{2} \right )} \right )}
          -
          \ln{\!\left ( \Gamma{\left ( \frac{1}{4} - i \frac{t}{2} \right )} \right )}
       \right \rbrack
      } 
\end{eqnarray}
which is exactly the phase of the Riemann zeta function. Two years after their
first result, Bhaduri {\emph{et al}}. extended this one-dimensional model to a
two-dimensional one where in one direction the potential is a traditional
confining parabolic potential, and in the perpendicular direction ($y$) they
kept the inverted harmonic oscillator~\cite{Bhaduri1997a}. This choice was
motivated by the analysis of the Gutzwiller trace formula on the $\sigma=1$
border of the critical line, and also by the form of the electrostatic potential
at the bottleneck of a quantum contact in a mesoscopic
structure~\cite{Buttiker1990}.

While the inverted oscillator reproduced the oscillating part of the
$\zeta(s)$ phase in Bhaduri's work, Berry and Keating showed that a
regularisation of a surprisingly simple one-dimensional classical
Hamiltonian, ${\cal{H}}=xp$, reproduces the smooth counting function
of the zeros~\cite{Berry1999}. We note here that this choice of
${\cal{H}}$ is a canonically rotated form of the inverted oscillator
Hamiltonian $\sim (p^{2} - x^{2})$. Moreover, the quantum mechanical
model of the corresponding symmetrised Hamiltonian, ${\cal{H}}=(xp +
px)/2$ has also been investigated and exactly solved preserving the
self-adjoint property of the Hamiltonian~\cite{Sierra2007, Twamley2006}.
The beauty of the $xp$- or inverted oscillator model is that
it satisfies most of the
properties listed earlier (see page \pageref{item:Berry1}): valid
as a classical mechanical model; the dynamics is one-dimensional and
uniformly unstable since the solution of the Hamiltonian equations
are exponentially decaying or diverging; it lacks time-reversal
symmetry. However, the trajectories are not bounded causing significant
hardship in the semiclassical quantisation. As in the hyperbolic case,
the boundary conditions or the way the phase space is regularised/
compactified become decisive. Berry and Keating suggested a simple regularisation
\cite{Berry1999} by introducing a cut-off in both position and momentum.
This process results in a finite area, which can be filled up with
Planck-cells of size $h$, thus counting the number of available quantum
states. Another approach is available if one notices the dilation
symmetry of the Hamiltonian ($x \mapsto \lambda x$, and $p \mapsto
p/\lambda$). This symmetry manifests itself in the transformation
of the wavefunction as
\begin{equation}
   \psi (\lambda x) = \frac{1}{\lambda^{1/2 - iE}} \psi(x)
\end{equation}
and one might suggest restricting ourselves to $\lambda$ being a
positive integer. This could be an attractive suggestion, because
the wave-packet, generated by the uniform superpositions of all
these transformed wavefunctions is
\begin{equation}
   \Psi (x)
   =
   \sum_{\lambda = 1}^{\infty}{\psi (\lambda x)}
   =
   \zeta \left ( \frac{1}{2} - iE \right )
   \psi(x).
\end{equation}
However, there is no physical motivation which would require this
$\zeta$ pre-factor to vanish. Furthermore this integer-based
dilation-symmetry does not form a group, because the multiplicative
inverse element (which would be $\lambda = 1/m$) is missing.

Berry and Keating also established a peculiar canonical transformation
($X = 2\pi/p$, $P = xp^{2}/2\pi$) for this Hamiltonian, which
exchanges and mixes the roles of the physical position and
momentum, but was uncertain ``how to convert this ``quantum
exchange'' into an effective boundary condition'' \cite{Berry1999a}. Aneva
also analyses this boundary condition for a hyperbolic dynamical
system with conformal geometry and shows how this exchange
transformation arises as a result of boundary conditions
\cite{Aneva1999, Aneva2001, Aneva2001a}.

Later Sierra generalised Berry's model in two different ways,
first by incorporating the fluctuation terms, $d_{\mathrm{osc}}(E)$
(see relation (\ref{eq:RZ_DensityOfZerosOscillatoryPart})), via changed
boundary conditions \cite{Sierra2008a}. Together with Townsend,
they also considered the motion of a charged particle (electron)
moving on the $[xy]$ plane in a constant uniform perpendicular
magnetic field, and in an electric potential described by the
following Hamiltonian
\begin{equation}
   {\cal{H}}
   =
   \frac{1}{2\mu}
   \left \lbrack
      p_{x}^{2}
      +
      \left ( p_{y} + \frac{eB}{c} x \right )^{2}
   \right \rbrack
   +
   e \lambda x y.
\end{equation}
In this model, the number of semiclassical quantum states with energy
less than $E$ has the same functional form as the counting function
of the $\zeta(s)$ zeros, eq.~\ref{eq:RZCountingFunctionSmoothPart},
i.e. the smooth part of the Riemann-zeros is reconstructed by the
lowest-lying Landau level of the charged particle. The fluctuation
term -- as they speculate -- might be explained by the contribution
of higher Landau-levels. This surmise, however, is only supported by
estimating the order of magnitude of these higher contribution and
comparing it to that of the Riemann $\zeta(s)$. This model has
the additional attraction of being potentially accessible to 
experimentalists, including in lower spatial dimensions \cite{Toet1991,
Li2007, Park2009}.

Exploiting the $x \leftrightarrow p$ exchange symmetry of this
model and using the Riemann-Siegel formula for the $\zeta(s)$
function, Sierra created a new model in which the Jost solutions
are directly proportional to the Riemann zeta function, and the
non-trivial zeros become the energies of the bound states. This
achievement does not, however, prove the Riemann Hypothesis, as
Sierra explicitly states ``we cannot exclude the existence of
zeros outside the critical line''.

In summary, we first introduced, motivated by Gutzwiller's
trace formula, a quantum mechanical model on a surface with negative
curvature, which lead us to the mathematically exact Selberg trace
formula. The importance of this result is at least twofold. Firstly,
it reassures us that describing chaotic systems via the periodic
orbits is likely to be feasible, and secondly demonstrates the role of
periodic orbits in a generic system in determining the smooth and
fluctuating parts of the density of states. We further elaborated on
another non-Euclidean model, proposed by Pavlov and Fadeev, in which
the Riemann-zeta function determines the $S$-matrix over the complex
energy plane.

Converting these results into the usual Euclidean space, however,
seems challenging. Although a few models have successfully reproduced
the smooth part of the density of quantum states, the fluctuation terms
of these models differ from that of the Riemann zeta function.

\subsubsection{\label{seq:qm_BoundStates}Bound state models}

From the 1950s a new approach, the Random Matrix Theory, emerged from the study
of the spectrum of heavy nuclei. The same statistical apparatus had also been
used to analyse the statistical properties of the seemingly random Riemann zeta
zeros, and lead to the conjecture that the $\zeta(s)$ zeros belong to one
particular universality class~\cite{Bohigas1984a, Bohigas1986}, the so-called
Gaussian Unitary Ensemble (see later in section \ref{sec:NuclearPhysics}). This
result suggested property \ref{item:Berry3} on Berry's list. However, Wu and
Sprung generated a one-dimensional, therefore integrable, quantum mechanical
model which can possess the Riemann zeta zeros as energy eigenvalues~\cite{Wu1993}
and show the same level-repulsion as that observed in quantum chaos. This was a
contradictory result since on one hand the $\zeta(s)$ zeros follow a statistics
specific for systems violating time reversal symmetry, on the other hand, Wu
and Sprung's model, by definition, was invariant under time reversal. However,
the proposed model was not lacking in irregularity, since the potential
reproducing the Riemann zeta zeros appeared to be a fractal, a self-similar
mathematical object. Nevertheless, these authors derived for the first time
a smooth, semi-classical potential which generate the smooth part of $N(E)$%
\footnote{In the mathematical literature the argument is usually denoted by
$T$, as in section \ref{sec:HistoryAndNecessities}. Motivated by physics,
we use here $E$.}, through
\begin{equation}
   N(E)
   =
   \frac{1}{h} \iint_{{\cal{H}} \le E}{\! dx\, dp}
   =
   \frac{2}{\pi}
   \int_{0}^{x_{\mathrm{max}}}%
       {\!\!\sqrt{E-V(x)}\, dx}
\end{equation}
with the $2m/\hbar^{2}$ set to unity. Solving this Abel-type integral equation
one may derive the following implicit expression for $V(x)$
\begin{equation}
   x(V)
   \! = \!
   \frac{1}{\pi}
   \!\left \lbrack
      \sqrt{V-V_{0}} \ln{\!\left ( \frac{V_{0}}{2\pi} \right )}
      +
      \sqrt{V}
      \ln{
          \!\left (
          \frac{\sqrt{V} + \sqrt{V-V_{0}}}%
               {\sqrt{V} - \sqrt{V-V_{0}}}
           \right )
         }
   \right \rbrack
\end{equation}
where $V_{0}$ has to be chosen such that the potential is not multi-valued, i.e.
$V_{0} \le 2\pi$. The choice of $V_{0}$ affects the potential at its bottom ($x
\approx 0$), but for large $x$ it does not have a significant impact and for $x
\gg 1$
\begin{equation}
   x(V) = \frac{\sqrt{V}}{\pi} \ln{\!\left ( \frac{2V}{\pi e^{2}} \right )}
\end{equation}
(see figure 1 in \cite{Wu1993}). We note here that Mussardo, using similar
semiclassical arguments as Wu and Sprung, recently also gave a simple expression
for a smooth potential supporting the prime numbers~\cite{Mussardo1997} as
energy eigenvalues. Furthermore, Mussardo also proposed a hypothetical resonance
experiment to carry out primality testing; this theoretically infinite potential
could be truncated at some high energy; thus transforming it into a finite
well. If an incident wave radiated onto this well has energy $E=n \hbar \omega$
where $n$ is a prime number, then it should cause a sharp resonance peak in the
transmission spectrum - argues Mussardo.

Turning back to the smooth potential studied by Wu and Sprung, which is
able to `roughly' reproduce $N(E)$, it is then modified to have the low
lying $\zeta(s)$ zeros exactly. In order to achieve this goal Wu and
Sprung set up a least-square minimisation routine, to minimise the
difference between the actual energy eigenvalues and the exact zeros.
The result was surprising, since the potential curve became coarse and
resembled a random potential. They analysed this curve using the standard
box-counting technique and measured a $d=1.5$ fractal dimension for the
potential reconstructing the Riemann zeta zeros. 

Ramani {\emph{et al}}. pointed out \cite{Ramani1995} that the apparent
contradiction between Berry's conjecture and Wu and Sprung's model, i.e.
whether or not the physical system exhibits time-reversal symmetry, is
caused by the coarse curve of the potential, since {\emph{any}}
smooth one-dimensional potential would lead to locally evenly spread energy
levels, which is not the case for Wu's potential. They also provided a very
efficient algorithm, the ``dressing transformation'' with which one can build up
the quantum potential from individual energy eigenvalues. However, they
standardised the spectrum using the `spectrum unfolding' technique which
eventually lead them to the conclusion: the fractal dimension of the potential
supporting the Riemann zeta zeros has $d \rightarrow 2$ rather than that
measured by Wu and Sprung. In a reply~\cite{Wu1995}, Wu and Sprung pointed out
that this difference in fractal dimension is putatively caused by the
alternative choice of spectrum. As they argued, Ramani's spectrum does not have
the same average density, long range correlation and nearest-level spacing
distribution as the Riemann zeta function, therefore one cannot draw valuable
conclusions regarding the potential.

\begin{figure}
   \includegraphics[angle=-90, width=0.47\textwidth]{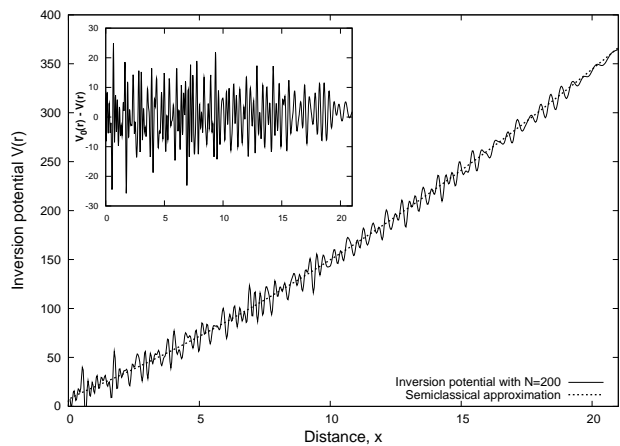}
   \caption{\label{fig:QM_RiemannZetaPotential}
            Main figure shows the semi-classical potential (dashed line), and
            the fractal potential (solid line) supporting the first two hundred
            zeros of $\zeta(s)$ as energy eigenvalues. The inset depicts the
            difference of these potentials. From \cite{Schumayer2008}.}
\end{figure}

Nearly a decade after Wu and Sprung's original article, van Zyl and Hutchinson
attempted to clarify the questions raised by the two previous
works~\cite{vanZyl2003}. They showed that for the same set of energy levels
different potential generating techniques (the variational approach used by Wu
and Sprung, the dressing-transformation used by Ramani {\emph{et al}}.) lead to
the same potential, depicted in Figure \ref{fig:QM_RiemannZetaPotential}. This
result had been further strengthened by Schumayer {\emph{et al}}. who used the
inverse scattering transformation as a third technique obtaining the same
potentials as in the earlier works~\cite{Schumayer2008}. It is noteworthy
to mention that the inverse scattering transform guarantees the uniqueness
of the potential in one-dimension. This analysis, therefore, elucidated
that the difference in measured fractal dimension cannot originate from the
method of inversion. Moreover, they confirmed $d=1.5$ for the Riemann zeta
potential. These works all demonstrated the importance of long-range
correlations in determining the fractal dimension of the potential.

In a similar manner to that of the the Riemann $\zeta(s)$ zeros, the prime
numbers can also be considered as an energy spectrum, thus a potential can
be associated with them and it also proves to be fractal, but with a larger
fractal dimension, $d=1.8$. This result is somewhat puzzling. The two sets,
those of the zeta zeros and the prime numbers can be mapped onto each other
via eq. (\ref{eq:ExactFormulaForPi}), but the nearest-neighbour spacing
distribution of prime numbers is known to be Poisson-like (almost uncorrelated
random distribution) while that of the Riemann zeros is rooted in the Gaussian
Unitary Ensemble, and exhibits the corresponding correlations (see expression
(\ref{eq:NP_MontgomeryPairCorrelation}) in section \ref{sec:NuclearPhysics}).
One may, therefore, conclude that Riemann's formulae converts two very different
random distributions into each other, or as Sakhr {\emph{et al}}. put it
\cite{Sakhr2003}: ``it is possible to generate the almost uncorrelated
sequence of the primes from the interference of the highly-correlated
Riemann zeros''.

Regarding the fractal nature, Schumayer {\emph{et al}}. also established that
the potentials associated with either the zeros of $\zeta(s)$ or with the prime
numbers are multi-fractals, i.e. these potential curves cannot be characterised
by one number $d$, but a range of dimension is necessary to describe their
properties (for definition see \cite{Schumayer2008}).

%

Finally, at the end of this section devoted to the quantum mechanical
models of the Riemann zeta function, we briefly refer to another
alternative spectral interpretation of the zeros proposed by Connes
\cite{Connes1999}. During the comparison of the Gutzwiller's trace formula
for quantum mechanical systems and that of the $\zeta(s)$ function we
noticed the overall sign difference in $d_{\mathrm{osc}}$ (see the
negative sign in equation (\ref{eq:RZ_DensityOfZerosOscillatoryPart})
in front of the summation), i.e. the contribution of the periodic orbits
should be subtracted and not added to the smooth density of states,
${\bar{d}}(T)$ \cite{Berry1986}. This sign difference led Connes to
interpret the zeros as gaps, missing lines from the otherwise continuous
energy spectrum rather than discrete energy levels.

\subsection{\label{sec:NuclearPhysics}Nuclear physics}

\noindent 
{\emph{Random Matrix Theory (RMT) has been successfully applied to predict
ensemble averages of observables for heavy nuclei. Even though the Riemann zeros
are distributed randomly, some of their statistical quantities correspond to
that of the Gauss Unitary Ensemble. We discuss the RMT briefly for historical
reasons. The reason for brevity owes to two recent Colloquia devoted to RMT
\cite{Papenbrock2007, Weidenmuller2009}.}}

Unfortunately the degrees of freedom of even a moderately large
nucleus are still far beyond our computational capability, be it
analytical or numerical. Similar problems, although the number of
components are on a different scale, have occurred before in
physics and engendered the development of a new branch of physics,
{\emph{statistical mechanics}}. This is exactly what Wigner had in
mind when he suggested a statistical description of nuclei
\cite{Wigner1951}. He suggested that nuclei can be statistically
described using random matrices carefully chosen from pre-determined
ensembles. The new description emerging from this examination is
the {\emph{Random Matrix Theory}}.

Although random matrix theory emerged from the statistical description
of nuclei, it has already infiltrated into many different areas of
physics. Recent developments of this branch of physics have been
reviewed in \cite{Bohigas1989, Forrester2003, Weidenmuller2009}.
Moreover we can suggest the monograph by one of the leading figures
of random matrix theory~\cite{Mehta2004}.

But how to choose the ensemble of random matrices suitable for a certain
system, or for the Riemann $\zeta(s)$ function? Throughout classical
mechanics symmetry plays a decisive role in determining the dynamics of
different systems. If a physical system has a symmetry it implies, via
Noether's theorem, the existence of a conserved quantity, e.g. the translational
invariance in time dictates energy conservation, continuous rotational
invariance requires the angular momentum remain constant. These symmetries
limit the possible forms of the Hamiltonian describing the given system.
Therefore, if one wants to approximate a Hamiltonian with a large, but finite
dimensional matrix these symmetries will determine the type and structure of
the matrix, whether it is real or complex, symmetric or
hermitian~\cite{Dyson1962}.

In the case of an integrable system, the conserved quantities are all known.
Therefore the Hamiltonian can be diagonalised, with each eigenvalue forming
its own symmetry-class. This leads to the assumption that these eigenvalues
are completely uncorrelated. Let us also assume that the average spacing
between eigenvalues is unity in the overall sequence of eigenvalues. If $p(s)$
denotes the probability distribution of nearest neighbour spacings, i.e. if
$\epsilon_{1}$ and $\epsilon_{2}$ are eigenvalues of the given system, then
$\epsilon_{1} - \epsilon_{2} = s$, then one can express~\cite{Stockmann1999}
the probability of finding two eigenvalues in a distance between $s$ and $s+ds$
with no other eigenvalues in between. Dividing the distance $s$ into $N$ equal
intervals, the probability is simply
\begin{equation}
   p(s) ds
   =
   \lim_{N \rightarrow \infty}%
       {\!\left \lbrack
           \left (
                 1 - \frac{s}{N}
           \right )^{N}
        \right \rbrack
       }
   ds\ .
\end{equation}
In the $N \rightarrow \infty$ limit the right hand side becomes the
exponential function. Therefore the probability distribution $p(s)
= \exp{\!(-s)}$ is the Poisson distribution with parameter equal to
1. This is quite a general result for integrable systems as Berry
and Tabor have demonstrated~\cite{Berry1977}. Similarly, one can
deduce similar probability distributions for {\emph{universality
classes}} of random matrices, e.g. Gaussian Unitary Ensemble, Gaussian
Orthogonal Ensemble, etc. The classification refers to the universality
conjecture: if the classical dynamics is integrable then $p(s)$
corresponds to the Poisson ensemble, while in the chaotic case
$p(s)$ coincides with the corresponding quantity for the eigenvalues
of a suitable ensemble of random matrices~\cite{Bohigas1984a}.
Furthermore, the local statistics of the eigenvalues converge
as the order of the matrix increases.

How is this connected to the Riemann-zeta zeros? The zeros can be
treated as eigenvalues of a fictitious physical system, just as
Hilbert and P{\'o}lya suggested, and their statistical properties
examined. In 1973 Hugh Montgomery showed \cite{Montgomery1973}
that the pair-correlation function of the zeros is
\begin{equation} \label{eq:NP_MontgomeryPairCorrelation}
   r_{2} (x) = 1 - \left ( \frac{\sin{(\pi x)}}{\pi x} \right )^{2},
\end{equation}
provided the Riemann Hypothesis is true. Freeman J. Dyson, during
an informal discussion over tea \cite{Cipra1999}, pointed out to
Montgomery that this is exactly the same result as obtained for
random matrices picked from the Gaussian Unitary Ensemble. However,
this statement is made in the asymptotic limit, i.e. as one goes to
infinity on the critical line, $1/2 + i E$, or in the RMT language
as the size of the matrices, $N$, tends to infinity. At finite height
$E$ or dimensionality $N$ discrepancies may occur compared to expression
(\ref{eq:NP_MontgomeryPairCorrelation}). Interestingly, it was shown
using heuristic arguments, that the nearest-neighbour spacing
distribution of the zeta zeros and that of unitary random matrices of
finite dimension are the same \cite{Bogomolny1995}. Moreover, the same
authors extended their study of correlation functions~\cite{Bogomolny1996},
$r_{n}$ of order $n$ ($n \ge 2$) and proved, in the appropriate
asymptotic limit, $r_{n}$ of the Riemann zeta zeros are equivalent
to the corresponding GUE result. This result was complimentary to
Montgomery's second order \cite{Montgomery1973}, Hejhal's third
order~\cite{Hejhal1994} and Rudnick and Sarnak's general result
for the $n$th order correlation function~\cite{Rudnick1996}.

What does this result demand from a model of the Riemann zeros?
The striking similarity between the 
pair-correlation function of the $\zeta(s)$ zeros and the eigenvalues
of random matrices from the GUE ensemble only holds for short-range
statistics. Odlyzko, by calculating the statistics for substantial
numbers of zeros, carried out an empirical test \cite{Odlyzko1987}
and confirmed Berry's predictions \cite{Berry1985} about the discrepancies
between the GUE theory and computed behaviour of the $\zeta(s)$ zeros.
The long-range correlation and the small spacing statistics of the
$\zeta(s)$ zeros noticeably deviate from the GUE prediction. This is
expected \cite{Berry1985, Berry1988}, since long-range correlations are dominated
by the short periodic orbits, which are system specific and therefore not
universal. For $\zeta(s)$ the mean separation between zeros is $\ln{\!(E/2\pi)}$ 
while the smallest period is $\sim \ln{\!(2)}$ (Table \ref{tab:Dictionary}).
Conclusively, the GUE-predicted universal correlation for zeros near $E$
should fail beyond $\ln{\!(E/2\pi)}/\ln{\!(2)}$ \cite{Berry1999a}.
Despite the deviation explained above, the statistics of the $\zeta(s)$
zeros asymptotically coincide with those of the GUE ensemble, consequently
the corresponding quantum system ought to violate time-reversal symmetry
\cite{Berry1999, Berry1999a}. This may have motivated Berry and Keating's
choice of a $\sim (xp+px)$ as a Hamiltonian. 

Finally, we must mention an unexpected spin-off result of random
matrix theory related to the P{\'o}lya-Hilbert conjecture. Crehan
asserted \cite{Crehan1995} that for any bounded sequence there are
infinitely many {\emph{classically}} integrable Hamiltonians for
which the corresponding quantum spectrum coincides with this
sequence. Furthermore, as an example for his theorem, he shows
that infinitely many classically integrable non-linear oscillators
are capable of exactly reproducing the Riemann-zeta zeros when they are
quantised. Unfortunately, the theorem is an existence theorem and
not a constructive one. If such a system could be created, whether
physically or just theoretically, that would be aesthetically pleasing:
it would connect the most studied physical model (oscillator) with
the basis of our arithmetic (prime numbers). Crehan's result is
promising and is also supported by the relationship between the
Riemann $\zeta(s)$ zeros and the Painlev{\'e} V equation, the latter
of which plays a central role in the theory of completely integrable
dynamical systems \cite{Ablowitz1991}.

Finally, in this section we briefly mention the notion of
{\emph{quantum ergodicity}} which attracted substantial attention
in the last three decades in the search for links between classical
and quantum ergodicity, i.e. what ``fingerprint'' the classical
chaos leaves in the physical properties if we quantise the system,
especially in the long-time behaviour. Only few rigorous results
\cite{Schnirelman1974, Verdiere1985, Zelditch1996} are known, and
one of them says that the expectation value of operators over
individual eigenstates is almost always the ergodic, microcanonical
average of the classical version of the operator. However, the
theoretically rigorous understanding of quantum ergodicity is still
in its infancy. Numerical simulations suggest though that quantum
chaotic systems exhibit universal behaviour at a particular length
scale, and at this scale the statistics of the eigenvalues resemble
that of large random matrices chosen from specific ensembles
\cite{Berry1977a, GutzwillerBook, Bohigas1984a, Bohigas1984b,
Heller1984, Agam1995}. It is unfortunate that this length scale
is so minute that it hinders the numerical simulations substantially.
Nevertheless, it has also been shown theoretically \cite{Tomsovic1991,
Kaplan1996} that quantum eigenstates must deviate from the RMT
predictions. These corrections may stand out from the spread out
background of RMT, just as the unstable periodic orbits do as eigenstates
with enhanced amplitudes as depicted in Figure \ref{fig:QM_ScarsInStadiumBilliard}.
Although further numerical simulations \cite{Backer1998, Kaplan1999}
provide some evidence regarding the connection between RMT and quantum
ergodicity, its interpretation and strength remain open questions.

\subsection{\label{sec:CondensedMatterPhysics}Condensed matter physics}

\noindent 
{\emph{In condensed matter physics the fundamental structure is the crystal
lattice. Below we examine the connection of the lattice with the generalised
Riemann hypothesis. We also show how the specific heat capacity of a solid
restricts the location of the $\zeta(s)$ zeros.}}

One of the fundamental bases of modern condensed matter physics is the
geometrical structure of solids; the lattice. The examination of this
mathematical structure is necessary to understand even the basic properties
of matter. The regular structure of a perfect lattice is suitable
for immediate comparison with regularities among the natural numbers,
and therefore it is not a surprise that many number-theoretical functions
arise in crystallography, e.g. Ninham {\emph{et al.}} present a witty review on the
M{\"o}bius function \cite{Ninham1992}. For those mathematically more
inclined we suggest the book ``From Number Theory to Physics'' by
Waldschmidt \cite{Waldschmidt1995}. Moreover, not only the perfect
regularity of a lattice, but also the lack of this regularity can
be related to the Riemann zeta function, as Dyson indicated recently
\cite{Dyson2009}: ``A fourth joke of nature is a similarity in behavior
between quasi-crystals and the zeros of the Riemann Zeta function.''
In the following, we briefly examine why a solid state constituted by
ions should even exist, what binds these ions to each other?

Ions arrange themselves into a structure which maximises the attractive
interaction between unlike and minimises the repulsive interaction between like
charges. In an ionic crystal, such as NaCl, the main contribution to the
binding energy has an electrostatic origin with the van der Waals term only a
few percent of the former. The electrostatic term is called the Madelung energy,
and the energy of one ion in the solid is called the Madelung constant.

For the sake of simplicity, let us first imagine a one-dimensional infinitely
long ionic lattice. Cations and anions are located next to each other at a
distance $a$, in a simplified NaCl structure. If simply two unit charges $q$
were positioned at the same distance $a$, the electric potential energy of one
of the charges would be ${\cal{U}} = q^{2}/4\pi \epsilon_{0} a$. In a solid each
ion is in the field of all the remaining charges, both positive and negative.
\begin{figure}[htb!]
   \includegraphics[width=0.46\textwidth]{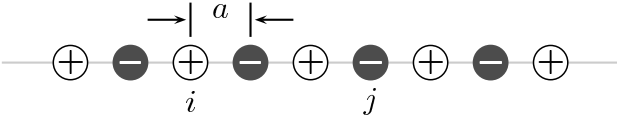}
   \caption{\label{fig:CM_1DLatticeMadelung}
            Schematic structure of a fictitious one-dimensional solid built up
            by cations and anions, positioned in alternating pattern.}
\end{figure}
The total electrostatic potential energy of one ion at position $i$ in the
lattice is therefore
\begin{equation} \label{eq:CM_TotalPotentialEnergyGeneral01}
   {\cal{U}}_{i}
   =
   \sum_{j \ne i}%
       {
        \frac{1}{4\pi \epsilon_{0}}
        \frac{(- 1)^{\abs{i-j}} q^{2}}{\abs{i-j} a}
       }
   =
   \frac{1}{4\pi \epsilon_{0}}
   \frac{q^{2}}{a}
   \sum_{k \ne 0}^{\infty}%
       {
        \frac{(-1)^{k}}{k}
       }
\end{equation}
where $j$ runs over all lattice sites except $i$ in the first summation, and in
the second we have changed the running variable to $k=\abs{i-j}$. In a finite
lattice we have $2 N$ ions, but in (\ref{eq:CM_TotalPotentialEnergyGeneral01})
each term belongs to two ions, therefore the total electrostatic potential
energy of the finite lattice is
\begin{equation}
   {\cal{U}}_{\mathrm{total}}
   =
   \frac{1}{2}\,
   2 N
   {\cal{U}}_{i}
   =
   N
   \frac{1}{4\pi \epsilon_{0}}
   \frac{q^{2}}{a}
   \sum_{k \ne 0}^{N}%
       {
        \frac{(-1)^{k}}{k}
       }.
\end{equation}
This form of ${\cal{U}}_{\mathrm{total}}$ can be divided into three terms: $N$
which guarantees the extensive nature of the energy, an energy factor, $q^{2}/%
4\pi \epsilon_{0} a$, and also a numerical factor depending only on the lattice
structure. One sees directly that the influence of the lattice on the total
electrostatic energy is comprised of an infinite sum. Since this energy term has
to be negative in order to describe binding, we incorporate this sign into the
Madelung constant $\alpha_{\mathrm{1D}}$ as
\begin{equation} \label{eq:CM_MadelungConstant1DSolid}
   \alpha_{\mathrm{1D}}
   =
   2
   \sum_{k = 1}^{\infty}%
       {
        \frac{(-1)^{k+1}}{k}
       }
\end{equation}
where the factor 2 appears because of the mirror-symmetry around the $i$th ion.
The total energy can be written as ${\cal{U}}_{\mathrm{total}} = -
\alpha_{\mathrm{1D}} N q^{2}/4\pi \epsilon_{0} a$, which is negative if
$\alpha_{\mathrm{1D}}>0$.

Generalising the NaCl structure we examined above for the realistic
three-dimensional case, one can write the Madelung constant for this
lattice as
\begin{equation} \label{eq:CM_MadelungConstant3DSolid}
   \alpha_{\mathrm{3D}}
   =
   \sum_{(i,k,l) \ne (0,0,0)}%
       {
        \frac{(-1)^{i+j+k+1}}{(i^{2} + j^{2} + k^{2})^{1/2}}
       }.
\end{equation}
Although it is tempting to evaluate this summation by approximating
the terms on concentric spheres centred at the reference ion ($i=j=k=0$)
and utilising the symmetry, the resulting series, $6 - 12/\sqrt{2}+8/
\sqrt{3}-$\dots is divergent which is physically unsatisfactory. The
convergence properties of such sums have been extensively investigated
\cite{Chaba1975, Chaba1976a, Chaba1976b, Chaba1977, Borwein1985}.
The sum (\ref{eq:CM_MadelungConstant3DSolid}) is an alternating and
conditionally convergent sum. The denominator of the summand is a
quadratic form, therefore the Madelung constant for a simple cubic
structure can be formally written as $\zeta_{\mathrm{EP}} (1/2,
\delta_{m,n})$ where $\zeta_{\mathrm{EP}}$ is the Epstein
zeta function (see below), and $m,n=$1, 2, 3. The second argument,
$\delta_{m,n}$, is determined by the type of the lattice, and in
crystallography it is a quadratic form ${\cal{P}}$ given by the
Gram matrix $p_{mn}={\mathbf{e}}_{m} {\mathbf{e}}_{n}$,
where ${\mathbf{e}}_{m}$ is the $m$th lattice vector. Therefore,
for example, the Madelung constant for the body-centered cubic
structure can be formally written as
\begin{equation}
   \alpha_{\mathrm{3D}}^{\mathrm{bcc}}
   =
   \zeta_{\mathrm{EP}} \!\left (
                                 1/2,
                                 \begin{pmatrix}
                                    2 & 1 & 1 \\
                                    1 & 2 & 1 \\
                                    1 & 1 & 2 \\
                                 \end{pmatrix}
                         \right )
   = 1.762675 .
\end{equation}
Here we have only dealt with the pure Coulomb-interaction, but this treatment
can be extended to screened electrostatic interactions as
well~\cite{Kanemitsu2005}.

The infinite sum in (\ref{eq:CM_MadelungConstant1DSolid}) strongly resembles the
Riemann-zeta function, except each term is weighted by a factor
$(-1)^{k+1}$, and its numerical value is $\alpha_{\mathrm{1D}} = 2 \ln{(2)}
\approx 1.3863$. Although in two and three dimensions the summation can be
written explicitly, obtaining a precise numerical value is far from easy and the
Epstein-zeta function is required. This function can be thought of as a generalised
zeta-function \cite{Ivic2003, Shanker2006} which is defined by
\begin{equation} \label{eq:Z_DefinitionOfEpsteinZeta}
   \zeta_{\mathrm{EP}} (s, {\cal{P}})
   =
   \sum_{{\cal{P}} \neq 0}%
       {
        \frac{1}{{\cal{P}}^{s}}
       }
\end{equation}
where ${\cal{P}}$ is a quadratic form defined on a $d$-dimensional lattice.
All lattice points for which ${\cal{P}} \equiv 0$ are excluded from the
summation. This function can be analytically continued to the same domain as
the Riemann-zeta function and also has its only pole at $s=1$ with residue
$\pi/ \sqrt{\Delta}$. The similarity goes further since $\zeta_{\mathrm{EP}}
(s, {\cal{P}})$ also satisfies a functional equation expressing mirror
symmetry. Thus, there is an inclination to generalise Riemann's conjecture:
all non-trivial zeros of $\zeta_{\mathrm{EP}} (s, {\cal{P}})$ have real part
one half. The temptation to do so is strengthened if one chooses specific
quadratic forms, e.g. $\zeta_{\mathrm{EP}} (s, {\mathrm{Id}}_{1}) = 2 \zeta(2s)$,
or $\zeta_{\mathrm{EP}} (s, {\mathrm{Id}}_{4}) \sim \zeta(s)\zeta(s-1)$, where
${\mathrm{Id}}_{n}$ is the $n$-dimensional identity matrix. Indeed, it was
shown eighty years ago that for binary quadratic forms (two dimensional
lattice), infinitely many zeros of $\zeta_{\mathrm{EP}}$ lie on the critical
line \cite{Potter1935} in a similar manner to Hardy for the Riemann zeta
function \cite{Hardy1914}. Remarkably, however, it has also been shown that
in any dimension one can construct such a ${\cal{P}}$, that the generalised
hypothesis {\emph{does not}} hold \cite{Terras1980}. This, admittedly negative,
result shows the intriguing connection between crystallography and this
generalised Riemann Hypothesis, but let us now depart from the abstract
and static crystal structure of solids, and examine the dynamics of this
system.

The lattice vibrations, phonons, are bosonic quasiparticles. Therefore if
one knows their energy spectrum, $\hbar \omega_{k}$, then the total energy of
the phonon gas is simply the sum over all modes of the crystal
\begin{equation}
   U
   =
   \sum_{k}%
       {
        \frac{\hbar \omega_{k}}{e^{\hbar \omega_{k}/k_{B}T}-1}
       }.
\end{equation}
Since the number of possible modes is large, $3N$, where $N$ is the number of
atoms in the lattice, one might convert this expression into an integral by
introducing the phonon density of states $g(\omega)$ normalised as $\int{
g(\omega)d\omega} = 3N$. Using standard methods to calculate the specific heat
of the solid, a directly measurable quantity, the following expression can be
obtained
\begin{equation}
   c_{\mathrm{V}}
   =
   \int_{0}^{\infty}%
       {
        \!\left (\! \frac{\hbar \omega}{k_{B}T} \right )^{2}
        \!\frac{e^{\hbar \omega/k_{B}T}}%
             {(e^{\hbar \omega/k_{B}T}-1)^{2}}
        \
        g(\omega) d\omega
       }.
\end{equation}
The only sample-specific quantity here is $g(\omega)$. Surprisingly, the number
theoretical M{\"o}bius function and the related M{\"o}bius inversion provides a
transformation to express $g(\omega)$ as a function of the measured specific
heat
\begin{equation}
   g(\omega)
   =
   \frac{1}{k_{B}\omega^{2}}
   \sum_{n=1}^{\infty}
       {
        \mu(n)
        {\cal{L}}^{-1}
        \!\left (
           \frac{c_{\mathrm{V}}(h/k_{B} u)}{u^{2}}
        \right )
       }
\end{equation}
where the inverse Laplace-transform, ${\cal{L}}^{-1}$, converts the space of
$u=h/k_{\mathrm{B}}T$ to $\omega/n$ \cite{Chen1990}. Around the same time another inversion
technique appeared in the literature~\cite{Xianxi1990} and was proven to be
equivalent to the one discussed above~\cite{Ming2003}. In the latter formulation
another special function, the Riemann zeta function was used, but in order to
avoid the dependence on the unproven Riemann Hypothesis a free ``regularisation''
parameter $s$ was also introduced. The density of states in this formalism
is
\begin{equation} \label{eq:CM_DaiDensityOfState}
   g(\omega)
   =
   \frac{1}{2\pi \omega}
   \int_{-\infty}^{\infty}
       {
       \frac{\omega^{ik+s} Q(k)}%
            {\Gamma(ik+s+2) \ \zeta(ik+s+1)}
       \
       dk
       }
\end{equation}
where $Q(k)=\int_{0}^{\infty}{u^{ik+s-1} c_{\mathrm{V}}(1/u) \ du}$.
Physically the density of states, $g(\omega)$ should be independent
of the regularisation parameter, although the existence of $Q(k)$
requires that $s$ must fall into the $0 \le s_{1} < s < s_{2}$ range
where $s_{1}$ and $s_{2}$ are the exponents of the specific heat
asymptotes at high- and low temperatures, respectively. Due to the
Dulong-Petit law, at high temperature the specific heat is independent
of the temperature, therefore $s_{1} \equiv 0$. On the other end of
the temperature scale the specific heat of phonons vanishes as $T^{d}$
in $d$ dimensions. Therefore $\zeta(s)$ in the denominator of the
integrand (\ref{eq:CM_DaiDensityOfState}) sweeps through the [$1$, $1+d$]
strip and ensures that no zeros of $\zeta(s)$ can occur there. Summarising,
the asymptotes of the specific heat contribution of lattice vibrations
in a solid provide an experimentally determined zero-free region of
$\zeta(s)$ on the complex $s$ plane. Although this offers no further
restriction than that which is already known from mathematics, it is
an example where physics places independent bounds upon the location
of the zeros.

\subsection{\label{sec:StatisticalPhysics}Statistical physics}

\noindent
{\emph{The description of both bosons and fermions relies on the mathematical
properties of the Riemann zeta function. We show how the problem of the `grand
canonical catastrophe' of number fluctuation in an ideal Bose-Einstein condensate
is connected to number theory. We introduce the concept of the primon gas,
and also consider number theoretical models of Brownian motion.}}

Although statistical physics, the physics of systems with a large number of
degrees of freedom, relied heavily upon combinatorics well before the birth of
quantum mechanics, probably the first appearance of the Riemann zeta function in
statistical physics occurred in Planck's momentous work on black body radiation, the dawn of
the quantum era. From then on, the Riemann zeta function pops up in numerous
different branches of statistical physics, from  Brownian motion to lattice gas
models.

Since the topic of ultra-cold quantum gases has expanded rapidly in the past
decade, we interpret the implications of the distribution of the Riemann zeta
zeros in this area first. We start with the non-relativistic, non-interacting,
spin zero Bose gas and treat the spatial dimension $D$ as a free parameter.
It is a standard textbook derivation~\cite{Huang2001} to show that this system
undergoes a phase transition at low temperatures, where the de-Broglie wavelength
$\Lambda$ of the particles becomes comparable to the inter-particle distance,
and thus the quantum nature of the constituents becomes decisive. Since the
particles are free, their spectrum is continuous, simply equal to the kinetic
energy $\epsilon= p^{2}/2m$. The total number of particles $N$ is the sum
of particles in each quantum state
\begin{eqnarray}
   N
   &=&
   \frac{1}{(2 \pi \hbar)^{D}}
   \int{f_{\mathrm{BE}}(\epsilon(p)) d^{D}q \ d^{D}p}
   \nonumber \\
   &=&
   \frac{V}{(2 \pi \hbar)^{D}}
   \int{
        \frac{d^{D}p}{e^{(\epsilon(p) - \mu)/kT} - 1}
       }
\end{eqnarray}
Changing the integration from momentum to energy leads directly to
\begin{equation} \label{eq:SP_BECInDDimensions}
   N
   \propto
   \int_{0}^{\infty}
       {
        \frac{\epsilon^{D/2-1}}%
             {e^{(\epsilon-\mu)/kT}-1}
        d\epsilon
       }
   \propto
   \zeta \!\left ( \frac{D}{2} \right )
\end{equation}
In the last step we used the fact that the chemical potential approaches the
energy of the lowest lying state, i.e. $\mu = 0$.

This result shows that the Bose-Einstein condensation phase transition cannot
occur in homogeneous non-interacting systems in dimensions lower than three. The
total number of atoms is a positive number and fixed for our system. In
one spatial dimension, since $\zeta(1/2)<0$, the positivity of $N$ cannot be fulfilled.
For two-dimensions the right hand side of (\ref{eq:SP_BECInDDimensions}) is
divergent due to the pole of the Riemann zeta function $\zeta(s)$ at $s=1$,
therefore $N$ appears to be infinite. The position of this pole can be
interpreted as the manifestation of the Mermin--Wagner--Hohenberg theorem, which
guarantees that a homogeneous two-dimensional system, provided the interaction
is sufficiently weak, cannot undergo a phase transition. One may thus see that
the pole structure of the Riemann zeta function determines whether our system of
interest can undergo a phase transition or not. We note here that this phase
transition can occur in lower dimensions for inhomogeneous
systems~\cite{Widom1968, Bagnato1991, Dai2003}. 

Let us turn to another fundamental question of statistical mechanics:
the equivalency of different statistical ensembles. The
difference between the predictions for the ``Riemann gas'' (see below) based on microcanonical,
canonical, and grand-canonical ensembles has been investigated by Tran and
Bhaduri~\cite{Tran2003}. However, the motivation for the analysis is rooted in
the so-called ``grand-canonical catastrophe'' of an ideal Bose
gas~\cite{Ziff1977}. The number fluctuation of an ideal boson gas is
\begin{equation}
   (\delta N)^{2}
   =
   \sum_{k=0}^{\infty}{\mean{n_{k}}\left ( \mean{n_{k}} + 1 \right )}
\end{equation}
where $\mean{n_{k}}$ denotes the ensemble average of the occupation number of
the $k$th energy eigenstate. According to the formula above, in the presence of
a macroscopically occupied ground state, the number fluctuation is proportional
to the total number of particles, $\delta N_{0} \sim N$, which, in the
thermodynamical limit ($N \rightarrow \infty$), leads to divergence.

Grossmann and Holthaus examined the illustrative model system of an ideal Bose
gas trapped in a $d$-dimensional potential with a power-law energy spectrum,
$\epsilon_{\nu_{i}} \sim \hbar \omega \nu_{i}^{\sigma}$, where $\nu_{i}$ labels
the energy eigenstates~\cite{Grossmann1997b, Weiss1997}. Later Eckhardt extended
the analysis to the mean density of states and the level spacing distribution for
ideal quantum gases \cite{Eckhardt1999}. Grossmann {\emph{et al.}} showed how the
dimensionality and $\sigma$, which, in some sense, measures the strength of the
potential, depress or enhance the number fluctuation of the ground state as a
function of the rescaled temperature $t = k_{B} T / \hbar \omega$:
\begin{equation*}
   (\delta N_{0})^{2} \sim
   \left \lbrace 
   \begin{matrix}
      C t^{d/\sigma}                            & & \hspace*{18pt} (0 < d/\sigma < 2) \\
      t^{2} \ln{\! ( t )}                       & & (d/\sigma = 2)     \\
      \zeta \! ( d/\sigma - 1  ) t^{2}          & & (2 < d/\sigma)     \\
   \end{matrix}
   \right .
\end{equation*}
where $C$ is calculated from a $d$-dimensional Epstein zeta
function~\cite{Holthaus2001}, although here its value does not play a
significant role. Therefore in a given spatial dimension the potential can
enhance the fluctuation while dimensionality depresses it. They also examined
the behaviour of the heat capacity around the critical temperature $t_{0}$ and
proved that the heat capacity changes continuously at $t_{0}$ if $1 <d/\sigma
\le 2$, but if $d/ \sigma > 2$ it undergoes a jump given by
\begin{equation} \label{eq:SM_HeatCapacity}
   \left . 
      \frac{C_{<} - C_{>}}{N k_{B}}
   \right \vert_{t_{0}}
   =
   \left (\frac{d}{\sigma} \right )^{2}
   \frac{\zeta \left ( \frac{d}{\sigma} \right )}%
        {\zeta \left ( \frac{d}{\sigma} - 1 \right )}
\end{equation}
where $C_{<}$ and $C_{>}$ denote the asymptotic values of the heat capacity at
$t \rightarrow t_{0}$ from below and above, respectively. It is
worthwhile to note that $(\delta N_{0})^{2}$ in the canonical ensemble could be
expressed as the following integral over the complex plane
\begin{equation} \label{eq:SM_NumberFluctuation}
   (\delta N_{0} )^{2}
   =
   \frac{1}{2\pi i}
   \int_{\tau - i \infty}^{\tau + i \infty}%
       {\Gamma(t) \Lambda(\beta,t) \zeta(t-1)}
\end{equation}
where $\Lambda(\beta, t) = \sum{(\beta \epsilon_{n})^{-t}}$ is the spectral
zeta function of a given spectrum $\epsilon_{n}$ and $\tau$ is chosen so all
the poles of the integrand lie on the left of the path of integration.
Therefore all the results shown above are determined by the pole structure
of the spectral and the Riemann zeta functions, $\Lambda(\beta,t)$ and $\zeta(s)$,
respectively. The large-system behavior is extracted from the leading pole,
while the finite-size corrections are encoded in the next-to-leading poles.

The formulae (\ref{eq:SM_HeatCapacity}) and (\ref{eq:SM_NumberFluctuation}) 
above did not just clarify an important physical question, namely number
fluctuation properties of a $d$-dimensional boson gas below the critical
temperature, but also had valuable number theoretical consequences. The
problem solved above is a purely combinatorial one~\cite{Grossmann1997a,
Holthaus2001, Weiss2002, Weiss2003}: how many ways can one distribute
$n$ excitation quanta over $N$ particles? This question, for general $n$
and $N$, is quite difficult. However, in the low temperature limit the
number of excitations, $n$, is much smaller than the number of particles,
$N$. This problem thus becomes tractable and one could obtain the results
mentioned above. Calculating the number fluctuation of a boson gas in a
one-dimensional ($d=1$) harmonic potential ($\sigma=1$) provides $(\delta
N_{0})^{2} \sim t$. But $t$ is simply proportional to the number of energy
quanta `stored' in the excited states, $t=(k_{\mathrm{B}}T/\hbar \omega)
= n$ and therefore $(\delta N_{0})^{2} \sim n$. A mathematician -- according
to Grossmann and Holthaus -- can now interpret this formula:
\begin{quotation}
	\noindent If one considers all unrestricted partitions of
	the integer $n$ into positive, integer summands, and asks
	for the root-mean-square fluctuation of the number of
	summands, then the answer is (asymptotically) just
	$\sqrt{n}$.
\end{quotation}
An intriguing consequence of this analysis is that a Bose-Einstein
condensate could be used (in theory at least) to factorise numbers
\cite{Weiss2004} which could be treated as a quantum computer
calculating the prime factors.

Furthermore, using their physical insight, Weiss and collaborators
could derive the following non-trivial number theoretical result.
Let $\Phi(n,M)$ denote the number of partitions of $n$ into $M$ summands
regardless of their order (e.g. $\Phi(5,2)=2$ while $\Phi(5,4)=1$),
and $\Omega(n)$ stand for the total number of different partitions,
i.e. $\Omega(n) = \sum_{m=1}^{n}{\Phi(n,m)}$. It is a natural step
to introduce the ``probability'' of having exactly $M$ terms in a
random partition by $p(n,M) = \Phi(n,M)/\Omega(n)$. It was then shown
that this probability distribution does not become Gaussian, and
it adopts its limiting distribution shape if $n> 10^{10}$, which
itself is a remarkable fact.

Here we only mention that the same combinatorial problem arises in many
different branches of mathematical physics, such as lattice animals in
statistical physics \cite{Wu1996, Lima2001}, numerical analysis on combinatorial
optimisation \cite{Mertens1998, Majumdar2002, Andreas2003, Bauke2003} and
also in the description of the low-energy excitations of a one dimensional
fermion-system as bosonic degrees of freedom ({\emph{bosonisation}})
\cite{Schonhammer1996}.

Tran and Bhaduri's, and then Holthaus and Weiss' works further
underline that the irregular behaviour of the canonical ensemble
lies in the combinatorics of partitioning integers and the
microcanonical and canonical ensembles prognosticate dramatically
different ground state number-fluctuations, $\delta n_{0}$. This
is an important example which unequivocally shows that the standard
statistical ensembles can not always be regarded as equivalent.

These examples, while not directly related to any attempt to
prove the Riemann Hypothesis, but rather just the zeta function,
do illustrate that results in physics can have profound
implications for mathematics in general and number theory in
particular.

The interpretation of prime numbers or the Riemann zeta zeros as energy
eigenvalues of particles appears not just in quantum mechanics but also
in statistical mechanics. Below, we review two concepts: the Riemann gas,
sometimes called the primon gas, and the Riemann liquid, although their
definitions vary slightly.

In 1990 Julia proposed the idea of a fictitious, non-interacting boson
gas~\cite{Julia1990}, where a single particle may have discrete energy equal to
$\epsilon_{0}$, $\epsilon_{1}$,\dots where $\epsilon_{n} = \epsilon_{0}
\ln{(p_{n})}$ ($n \ge 1$) and $p_{n}$ stands for the $n$th prime number. This is
why the constituents are called primons. Since the particles are not interacting,
a many-body state, in the second quantised formalism, can be represented by an
integer number $n$. This natural number has a unique factorisation,
$n=p_{1}^{m_{1}} p_{2}^{m_{2}} \cdots p_{k}^{m_{k}}$ which tells us that $m_{1}$
particles are in the $\vert p_{1} \rangle$ state, $m_{2}$ particles are in the
$\vert p_{2} \rangle$ state and so on. Due to this uniqueness, each many-body
state is enumerated once and only once. Therefore, the total energy of the
system, in the state $\vert n \rangle$, is $E_{n} = m_{1} \epsilon_{0}
\ln{\!(p_{1})} + m_{2} \epsilon_{0} \ln{\!(p_{2})} + \cdots + m_{k} \epsilon_{0}
\ln{\!(p_{k})} =  \epsilon_{0} \ln{\!(p_{1}^{m_{1}} p_{2}^{m_{2}} \cdots
p_{k}^{m_{k}})} = \epsilon_{0} \ln{\!(n)}$. In order to describe this gas we
have to construct the partition function from this spectrum
\begin{equation}
   {\cal{Z}}_{\mathrm{B}}
   =
   \sum_{n=1}^{\infty}%
       {\exp{\!\left (-\frac{E_{n}}{k_{B}T} \right )}}
   =
   \sum_{n=1}^{\infty}%
       {\frac{1}{n^{s}}}
   =
   \zeta(s)
\end{equation}
where $s=\epsilon_{0}/k_{B}T = \beta \epsilon_{0}$ and $\beta=(k_{B}T)^{-1}$ is
the inverse temperature. The partition function for the primon gas is thus the
Riemann zeta function $\zeta(s)$ and hence the alternative nomenclature. It is
apparent, by looking at the domain of $\zeta(s)$, that ${\cal{Z}}_{\mathrm{B}}$
is well-behaving for $s > 1$, i.e. at low-temperatures, while $s \le 1$ is
physically unacceptable. The boundary, $s=1$ represents a critical temperature,
called the Hagedorn temperature \cite{Hagedorn1965} above which the system cannot
be heated up, since its energy becomes infinite
\begin{equation}
   \mean{E}
   =
   -\frac{\partial}{\partial \beta} \ln{\!({\cal{Z}}_{\mathrm{B}})}
   =
   - \frac{\epsilon_{0}}{\zeta{(\beta \epsilon_{0})}}
     \frac{\partial \zeta(\beta \epsilon_{0})}{\partial \beta}
   \approx
   \frac{\epsilon_{0}}{s-1}.
\end{equation}
A similar treatment can be built up for fermions rather than bosons, but here
the Pauli exclusion principle has to be taken into account, i.e. two primons
cannot occupy the same single particle state. Therefore $m_{i}$ can be 0 or 1
for all $i$. As a consequence, the many-body states are labeled not by the
natural numbers, but by the square-free numbers. These numbers are sieved
from the natural numbers by the M{\"o}bius function. The calculation is a bit
more complex, but the partition function for a non-interacting fermion primon
gas reduces to the relatively simple form
\begin{equation} \label{eq:SM_FermionRiemannGas}
   {\cal{Z}}_{\mathrm{F}} = \frac{\zeta(s)}{\zeta(2s)}.
\end{equation}

The canonical ensemble is of course not the only ensemble used in statistical physics.
Julia extends the study to the grand canonical ensemble by introducing a
chemical potential $\mu$~\cite{Julia1994}, therefore replacing the primes $p$
with new `primes' $p e^{-\mu}$. This generalisation of the Riemann gas is called
the Beurling gas, after the Swedish mathematician Beurling who
generalised the notion of prime numbers. Examining a boson
primon gas with fugacity $-1$ shows that its partition function is
${\cal{Z}}_{\mathrm{B}}' = \zeta(2s)/ \zeta(s)$.

This last result has an astonishing interpretation. We know that for a system,
formed by two sub-systems not interacting with each other, the overall partition
function is simply the product of the individual partition functions of the
subsystems. Equation (\ref{eq:SM_FermionRiemannGas}) has precisely this
structure, there are two decoupled systems: a fermionic ``ghost'' Riemann gas at
zero chemical potential and a boson Riemann gas with energy-levels $E_{n} = 2
\epsilon_{0} \ln{\!(p_{n})}$. 

Julia also calculates the appropriate Hagedorn temperatures and analyses how the
partition functions of two different number theoretical gases, the Riemann gas
and the `log gas' behave around the Hagedorn temperature \cite{Julia1994}.
Although the divergence of the partition function signals the
breakdown of the canonical ensemble, Julia also claims that the continuation across
or around this critical temperature can help understand certain phase
transitions in string theory~\cite{Deo1989} or in the study of quark
confinement~\cite{Julia1994}. The Riemann gas, as a mathematically tractable
model, has been followed with much attention because the asymptotic density of
states grows exponentially, $d(E) \sim e^E$, just as in string theory. Moreover,
using arithmetic functions it is not extremely hard to define a transition
between bosons and fermions by introducing an extra parameter, $\kappa$ which
defines an imaginary particle, the non-interacting parafermions of order
$\kappa$. This extra parameter counts how many parafermions can occupy the same
state, i.e. the occupation number of any state falls into the $[0, \kappa-1]$
range, thus $\kappa=2$ belongs to normal fermions, while $\kappa \rightarrow
\infty$ represents normal bosons. The partition function of a free,
non-interacting $\kappa$-parafermion gas can be shown to be~\cite{Bakas1991}
\begin{equation}
   {\cal{Z}}_{\kappa} (s) = \frac{\zeta(s)}{\zeta(\kappa s)}.
\end{equation}
Bakas further demonstrates, using the Dirichlet convolution ($\star$), how one
can introduce free mixing of parafermions with different orders which do not
interact with each other
\begin{equation}
   f \star g = \sum_{d \vert n}{f(d) g \!\left ( \frac{n}{d} \right )}.
\end{equation}
where the shorthand notation $d \vert n$ means $d$ is a divisor of $n$.
This operation preserves the multiplicative property of the classically defined
partition functions: ${\cal{Z}}_{\kappa_{1} \star \kappa_{2}} =
{\cal{Z}}_{\kappa_{1}} {\cal{Z}}_{\kappa_{2}}$. It is even more intriguing how
interaction can be incorporated into the mixing by modifying the Dirichlet
convolution with a kernel function or twisting factor
\begin{equation}
   f \bullet g
   =
   \sum_{d \vert n}{f(d) g \!\left ( \frac{n}{d} \right ) K(n,d)}.
\end{equation}
Using the unitary convolution Bakas establishes a pedagogically illuminating
case, the mixing of two identical boson Riemann gases. He shows that
\begin{equation}
   ({\cal{Z}}_{\infty} \circ {\cal{Z}}_{\infty})
   =
   \frac{\zeta^{2}(s)}{\zeta(2s)}
   =
   \frac{\zeta(s)}{\zeta(2s)} \zeta(s)
   =
   {\cal{Z}}_{2} {\cal{Z}}_{\infty}
\end{equation}
Thus mixing two identical boson Riemann gases interacting with each other
through the unitary twisting, is equivalent to mixing a fermion Riemann gas with
a boson Riemann gas which do not interact with each other. This leads to the
interpretation that one of the original boson components suffers a transmutation
into a fermion gas. It is noteworthy to mention that the M{\"o}bius function,
which is the identity function with respect to the $\star$ operation (i.e. free
mixing) reappears in supersymmetric quantum field theories as a possible
representation of the $(-1)^{F}$ operator, where $F$ is the fermion number
operator~\cite{Spector1989, Spector1990, Spector1998}. In this context, the fact
that $\mu(n)=0$ for square-free numbers is the manifestation of the Pauli
exclusion principle.

It is therefore interesting that, what initiated as rather academic
studies to investigate potential attacks on the Riemann Hypothesis,
may lead to advances in physics. But let us return to the Hypothesis
through a slightly different definition of the Riemann gas. Here the
energy of the ground state is taken to be zero and the energy spectrum
of the excited state is $\epsilon_{n}=\ln{\!(p_{n})}$, where $p_{n}$
($n=$2, 3, 5, \dots) runs over the prime numbers. Let $N$ and $E$ denote
the number of particles in the ground state and the total energy of the
system, respectively. As we demonstrated above, the fundamental theorem
of arithmetic allows only one excited state configuration for a given
$E=\ln{\!(n)}$ ($n$ is an integer). It immediately means that this gas
preserves its quantum nature at any temperature, since only {\emph{one}}
quantum state is permitted to be occupied. The number fluctuation of
{\emph{any state}} (the ground state included) is therefore zero. In
contrast, the $\delta n_{0}$ predicted by the canonical ensemble is a
smooth non-vanishing function of the temperature, while the grand-canonical
ensemble still exhibits a divergence. This discrepancy between the
microcanonical (combinatorial) and the other two ensembles remains
even in the thermodynamic limit.

One may argue that the Riemann gas is fictitious and its spectrum is
unrealisable. However, the spectrum $\epsilon_{n} = \ln{\!( n )}$ does not
increase with $n$ more rapidly than $n^{2}$, therefore the existence of a
quantum mechanical potential supporting this spectrum is possible (cf. inverse
scattering transform used in section \ref{sec:QuantumMechanics}). The potential
has been given in \cite{Weiss2004}:
\begin{equation}
   V(x) = V_{0} \ln{\!\left ( \frac{\abs{x}}{L} \right )}
\end{equation}
where $V_{0}$ and $L$ are positive constants. Within the semiclassical
approximation the spectrum of this potential is
\begin{equation}
   \epsilon_{n}
   =
   V_{0} \ln{\!( 2n + 1 )}
   +
   V_{0} \ln{\! \left ( 
                        \frac{\hbar}{2L} 
                        \sqrt{ \frac{\pi}{2mV_{0}}}
                \right )
            }
\end{equation}
where $n=0$, 1,\dots and the second term only represents a constant energy
shift.

Recently, LeClair published two works~\cite{LeClair2007, LeClair2008} developing
and applying a finite-temperature field theoretical formalism for both boson and
fermion gases in low spatial dimensions in which he efficiently disentangles
zero temperature dynamics and quantum statistical sums for both the relativistic
and non-relativistic cases. His alternative approach is based on an $S$-matrix
formulation of statistical mechanics~\cite{Dashen1969}, which redefines the
quantum statistical mechanics directly in terms of dynamical filling fractions,
$f({\mathbf{k}})$. Assuming the two-body scattering kernel, ${\mathbf{K}}$, is
constant (i.e. constant scattering length) he derives, as pedagogical examples,
the well-known results for the boson
\begin{equation}
   T_{\mathrm{c}}
   \sim
   \left ( 
           \frac{n}%
                {\zeta\!(d/2)}
   \right )^{2/d}
\end{equation}
and also for the fermion gas
\begin{equation}
   \epsilon_{\mathrm{F}}
   \sim
   \left \lbrack
      \Gamma \left ( \frac{d+2}{2} \right ) n
   \right \rbrack^{2/d}.
\end{equation}
In two dimensions the critical temperature for the boson gas vanishes because of
the $\zeta(s)$ divergence at $s=1$, therefore this dimension needs further
consideration. Due to this instability, LeClair extends the examination for
energy-dependent two-body kernels, ${\mathbf{K}}=-\Re{\left ( \gamma_{\nu}
k^{2\nu-1} \right )}$ ($\nu$ is a complex number and $\gamma_{\nu}$ is constant),
for a one-dimensional fermion gas and explicitly constructs a quasi-periodic
potential, $V(x) \sim \cos{\!(\log{(x)})}/x^{2\sigma}$, in the real space which
reproduces the given kernel ${\mathbf{K}}$ in the two-body scattering
approximation. Furthermore, the thermodynamic variables, such as density and
pressure, are also shown to be physically valid (i.e. positive and have finite
value) provided $1/2<\Re{(\nu)}<3/2$. This fully covers the right hand side of
the critical strip divided by the critical line, and due to the symmetry of
$\zeta(s)$ this half-strip can be extended to the whole critical strip. His
argumentation is based on both the non-vanishing, non-divergent nature of the
physical quantities and also on the assumption that an interaction necessarily
modifies the thermodynamical quantities. If $\zeta(\nu)$ would be zero somewhere
in the critical strip, but off the critical line, then the leading order
contribution to the thermodynamical quantities would not be zero contradicting
the original assumption -- LeClair argues. This contradiction led him to
conclude that $\zeta(\nu)$ must be non-zero in the $1/2 < \nu < 3/2$ strip,
which can automatically be extended to the whole critical strip by using the
symmetries of the Riemann zeta function. LeClair, therefore, claims:
$\zeta(\nu)$ can have no zeros in the given range, consequently the
Riemann Hypothesis must be true. The basis for this conclusion however,
is itself an assumption and so does not constitute a proof of the Riemann
Hypothesis, but does provide another point of attack.

Examination of a similar fictitious, fermionic, many-body system
has also been considered by Leboeuf and lead to the conclusion
that ``time-periodic dynamical evolutions have to be considered
as serious candidates [for the Hilbert-P{\'{o}}lya
Hamiltonian]''~\cite{Leboeuf2001}.

At the end of this section, let us mention an interesting interlocking area of statistical
physics and number theory. A few authors have focused on the connection between
number theoretical functions and Brownian motion~\cite{Good1968, Billingsley1973,
Shlesinger1986, Wolf1998, Evangelou2005} or percolation~\cite{Vardi1998}. The
connection seems to be suggestive, especially if one defines the random motion
through the M{\"o}bius function $\mu(n)$, i.e. if $\mu(n)=\pm 1$ the particle
moves up- or downwards, and if $\mu(n)=0$ it does not move. Therefore the
distance of the particle from the origin after $n$ steps is $M(n) = \sum_{k}^{n}
{\mu(k)}$. The importance of this kind of Brownian motion lies in the so-called
Mertens conjecture. This states if $\abs{M(n)} \le \sqrt{n}$ then the Riemann
Hypothesis is true.

\begin{figure}
   \caption{\label{fig:SP_CumulativeSumOfMobius}%
            Function $M(n)$, the cumulative sum of the M{\"o}bius function is
            shown with the mean displacement of a random walk,
            $\sim \sqrt{n}$ for comparison.
           }
   \includegraphics[angle=-90, width=0.46\textwidth]{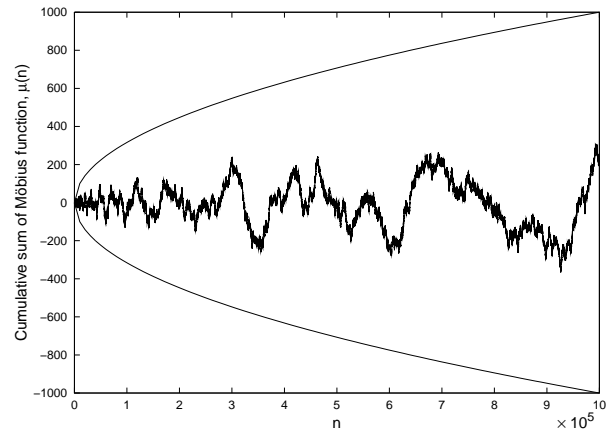}
\end{figure}

Figure \ref{fig:SP_CumulativeSumOfMobius} shows the path of the
particle for the first million steps. Although it is tempting to
conclude: the cumulative sum of $\mu(n)$ remains bounded by $\pm
\sqrt{n}$, this conjecture would actually be wrong as te Riele
and Odlyzko indirectly proved~\cite{Titchmarsh2003}. There is no
explicit counterexample known, but we have a loose interval [$10^{14},
\sim\! 3.6 \times 10^{10^{40}}$] in which there exists an $n$ such
that $M(n)/\sqrt{n} > 1$ \cite{Kotnik2006}. Nevertheless, the Mertens
conjecture is a sufficiency condition for the Riemann Hypothesis to
be true, not a necessary one. Its falsity therefore cannot invalidate
the Riemann Hypothesis. The failure of the Mertens Conjecture at such
a high $n$ value, however, does give cause for concern regarding
numerical evidence for the validity of the Riemann Hypothesis.

However, this is not the only possibility to define a random walk either
on the $\zeta(s)$ zeros or on the prime numbers. In the early 1970's
Billingsley defined a random, but finite, walk~\cite{Billingsley1973}
based on the fundamental theorem of arithmetic. 

Let $f(n)$ denote the number of prime factors of $n$ not counting their
multiplicity, e.g. $f(40)=2$, since $40=2^{3} \times 5$. It can be shown, that
on average, numbers below $N$ have $\ln{\!(\ln{\!(N)})}$ factors; a result which
on its own is a surprise. For example, numbers below $e^{e^{10}} \approx
10^{9566}$ have only 10 distinct factors on average. Based on the factorisation
one can define the following random walk: chose an integer $n \in [0,N]$,
starting from the origin we go up by a unit if 2 divides $n$ and down if it does
not, and continue the test with 3, 4\dots\ Although this construction does not
seem to be as random as a coin-tossing random walk and has few flows (e.g. it is
biased), Billingsley suggests a remedy to these problems and shows how the
similarity to Brownian motion leads to an Erd{\H{o}}s--Kac central limit
theorem for $f(n)$
\begin{equation}
   {\mathbf{P}}\!
   \left (
      \alpha \le
      \frac{f(n) - \ln{\!(\ln{\!(N)})}}%
           {\sqrt{\ln{\!(\ln{\!(N)})}}}
      \le \beta
   \right )
   \rightarrow
   \frac{1}{\sqrt{2 \pi}}
   \int_{\alpha}^{\beta}
        {e^{-u^{2}/2} \ du}.
\end{equation}
Therefore, the probability of $f(n)$ not deviating from the expected value
$\ln{\!(\ln{\!(N)})}$ more than $\alpha$ or $\beta$ times the standard deviation
can be estimated by a Gaussian integral. Therefore, the mapping of the number
theoretical problem onto a Brownian motion helps to derive a limit theorem for
the number theoretical function $f(n)$. As an example, if we chose $\alpha=-1$
and $\beta=1$ for $N=10^{9566}$ gives ${\mathbf{P}} \left ( -1 \le (f(n)-10)/
\sqrt{10} \le 1 \right ) \approx 0.68$, thus approximately 70\% of the numbers
below the chosen $N$ have from 6 to 13 distinct prime factors.

M. Wolf defined random walks in a different way~\cite{Wolf1998} and could
examine the distribution and correlation of twin-primes (where $p$ and $p+2$ are
both primes) and also of cousin primes ($p$ and $p+4$ are both primes). He also
suggested new random number generators with {\emph{theoretically}} infinite
period based on this kind of random walk, contrary to the widely used random
number generators~\cite{Press2007}. He also argues and with computations
demonstrates the multi-fractal nature of a subset of prime
numbers~\cite{Wolf1989}.

\section{Conclusion}

\hfill \parbox{80mm}{\raggedleft
                     {\emph{`All results of the profoundest mathematical
                            investigation must ultimately be expressible \\
                            in the simple form of properties of the integers.'}}\\
                     (Leopold Kronecker)
                     } \\
\vspace*{3mm}

Since this review is a summary itself in some sense, here we only attempt to
conclude with some general remarks.

In many respect the history of the Riemann Hypothesis is very similar to that
of Fermat's Last Theorem, which was stated in the seventeenth century and
solved 358 years later~\cite{Aczel1997, Ribenboim1999}, and along the path
towards the final proof it inspired and gave birth to new areas of mathematics,
such as the theory of elliptic curves. Although the Riemann Hypothesis has not
been proven or disproven it has already stimulated and influenced many areas of
mathematics, e.g. $L$-functions, which can be thought of as generalised zeta
functions, for which a generalised Riemann Hypothesis may hold. Interestingly,
for $L$-functions defined over functional space rather than the number field,
the similar hypothesis is rigorously proven.

Further evidence also suggests the validity of the Riemann Hypothesis,
let us just think of Levinson's theorem guaranteeing that at least one
third of the zeros are on the critical line. We, however, cannot
exclude the possibility of the existence of a counterexample to the
Riemann Hypothesis, i.e. a very high lying zero $s=\sigma + it$ for
which $\sigma \ne 1/2$. Similarly to the Mertens conjecture, the
counterexample may occur so high on the critical line, that we have
no machinery to even calculate zeros at that elevation. The immediate
impact of such a collapse of the Riemann Hypothesis would be immense
since there exist numerous ``proofs'' that are contingent upon it
\cite{Titchmarsh2003}.

That said, we cannot miss out in this review one computational masterpiece.
Not long after World War II, in which mechanical and electrical `computers'
were often used for encrypting messages (Enigma) and also for research (ENIAC),
Balthasar van der Pol constructed an electro-mechanical machine
which could calculate the first few zeros of the Riemann zeta
function~\cite{Pol1947}. This construction, despite its limited achievement,
deserves to be treated as a gem in the history of the natural sciences. Several
decades later, on the other end of the spectrum, a state-of-the-art application
of numerical techniques carried out by Brent, van de Lune, te Riele and Winter
\cite{Brent1979, Brent1982, Lune1983, Lune1986} calculated the first $1.5
\times 10^{9}$ zeros. Meanwhile Odlyzko~\cite{Odlyzko1987} explored the zeros
located around $t \sim 10^{20}$, and showed that all zeros (millions of them)
he found do exactly lie on the critical line. Here we note, that these numerical
checking are, of their own right, significant achievements, and also have
influenced the development of fast numerical techniques used in physics (see
e.g. \cite{Draghicescu1994, Greengard1994}).

In this review article we collected a few examples from different areas of
mathematical physics, starting with classical mechanics and finishing with
statistical mechanics, where the Riemann zeta function $\zeta(s)$, especially
its zero- and pole-structure, has a highly influential role.

In the section devoted to classical mechanics, we showed how the Riemann
Hypothesis can arise in a simple mechanical system, a ball bouncing on a rigid
wall. We also argued how these billiard systems lead to a revolutionary new way
of describing the dynamics of a chaotic system by introducing the evolutionary
operator. Here we also sketched the connection, a trace formula, between the
dynamics of a chaotic system and the periodic orbits of the same system.

This new descriptive language of dynamics through the trace formulae of the
Green function is suitable to develop a new quantisation technique for chaotic
quantum systems which otherwise was impossible using the standard Bohr
quantisation rules. Gutzwiller's trace formula has been explicitly mentioned,
because the Riemann zeta function obeys a very similar expression. Therefore, we
could compare the two formulae, (\ref{eq:RZ_DensityOfZerosOscillatoryPart}) and
(\ref{eq:QMGutzwillerTraceFormula_dosc}), and imagine what properties a quantum
system might have if its spectrum mimicked the zeros of the Riemann zeta
function.

We also surveyed two other attempts to find a quantum system which has a
connection to the Riemann zeta function. Both of these directions try to
associate $\zeta(s)$ with the spectrum of the system. The difference between
these approaches is that one of them relates $\zeta(s)$ to the positive energy
spectrum, i.e. scattering states, while the other, based on the
Hilbert-P{\'o}lya conjecture, proposes systems where the negative energies, thus
the bound states of the system, coincide with the zeros of $\zeta(s)$. This
latter case naturally guides us to condensed matter physics and statistical
mechanics, where one has to evaluate physical observable on the lattice points,
or derive all thermodynamical properties of a given particle-system provided the
spectrum is given.

In the sections concentrating on condensed matter physics, we first
showed how the Riemann zeta function, or one of its ancillary functions,
arose when we calculated the binding energy of a given structure of
solid matter. Finally we showed how physical requirements for the
specific heat of a solid can provide zero-free regions for the Riemann
zeta function. Research in this direction eventually may offer narrower
zero-free regions, and complement the approach in pure mathematics.

In the last section, we discussed three main areas of statistical physics where
the Riemann zeta function and its number theoretical aspects influence the
behaviour of a physical system. Firstly, we considered the low-temperature phase
transition of bosons, and showed that the pole structure of $\zeta(s)$
prohibits Bose-Einstein condensation in one- and two-dimensional uniform
systems. We also reviewed the `grand canonical catastrophe' of an ideal
Bose gas, where the predictions of two ensembles widely used in statistical physics
contradict each other, showing, therefore, that these ensembles cannot be
equivalent to one another. Finally, we examined a possible Brownian motion model
for the number theoretical M{\"o}bius function.

It would not be without precedent if a completely new theory
or a new mathematical language is needed in which the Riemann
Hypothesis can be `worded' naturally for the hypothesis to be
finally proved. As has happened earlier with mathematics,
natural science, and in particular physics, can give impetus
and  motivate new directions perhaps leading to the final proof.
It is amazing and captivating to see that a purely number
theoretical function has so many direct links to classical
and modern physics.

Nowadays we are not surprised by Galileo's famous keynote: ''{\emph{[Nature] is
written in the language of mathematics, and its characters are triangles,
circles, and other geometric figures without which it is humanly impossible to
understand a single word of it; without these, one wanders about in a dark
labyrinth}}''~\cite{Drake1957}. Probably we are not meandering in a labyrinth,
but we are definitely puzzled by the overwhelming difficulty of proving the
Riemann Hypothesis. We simply do not know as yet whether 
physics will ultimately help in understanding such an elegant mathematical statement as the
Riemann Hypothesis, but we are definitely witnessing the intertwining and
invigoration of both disciplines. The authors can only express their
hope that this work has to some extent captured the imagination of the Reader and, if so, it
has fulfilled its intended aim.

\section*{Acknowledgments}
Daniel Schumayer expresses his gratitude to Brandon P. van Zyl
for creating an inspiring and welcoming environment and partial
financial support through a grant from the National Science and
Engineering Research Council of Canada. We would also like to
sincerely thank all the referees of this manuscript. Their
comments and suggestions were highly valuable and lead to
a significant increase in the final quality of this paper.

This work has been financially supported by the University of
Otago and the Government of New Zealand through the Foundation
for Research, Science and Technology under New Economy Research
Fund Contract No. NERF-UOOX0703.


\bibliographystyle{apsrmp4-1}

\end{document}